\documentclass[12pt,a4paper,english,superscriptaddress,aps,nofootinbib]{revtex4}
\usepackage[utf8]{inputenc}
\usepackage[T1]{fontenc}
\usepackage{amsmath,amssymb,graphicx}
\makeatletter
\usepackage{babel}
\usepackage[active]{srcltx}
\usepackage{graphicx,color}
\usepackage{subfigure}
\usepackage{changebar}

\usepackage{hyperref}
\hypersetup{colorlinks=true,urlcolor= blue,citecolor=blue,linkcolor= blue}

\usepackage{braket}
\usepackage{dsfont}
\usepackage{mathtools}
\usepackage{slashed}
\usepackage{empheq}
\usepackage{tikz}
\usepackage{multirow}
\usetikzlibrary{decorations.pathmorphing}
\usepackage{graphicx}
\usepackage{amsfonts}
\usepackage{amsmath}
\newcommand{\revision}{\textcolor{black}}

\newcommand{\be}{\begin{equation}}
	\newcommand{\ee}{\end{equation}}
\newcommand{\bea}{\begin{eqnarray}}
	\newcommand{\eea}{\end{eqnarray}}

\begin{document}

 %\title{Smoothing of the translational symmetry in the $\phi^4$ theory}
 
\title{\revision{Soft breaking of the $\mathbb{Z}_2$ symmetry in the $\phi^4$ theory}}

\author{F. C. E. Lima}
\email[]{E-mail: cleiton.estevao@fisica.ufc.br (Corresponding author)}
\affiliation{Centro de Matématica, Computação e Cognição (CMCC), Universidade Federal do ABC (UFABC), Av. dos Estados 5001, CEP 09210-580, Santo Andr\'{e}, S.P. Brazil.}

\author{C. A. S. Almeida}
\affiliation{Departamento de F\'{i}sica, Universidade Federal do Cear\'{a} (UFC), Campus do Pici, Fortaleza - CE, 60455-760, Brazil.}

\begin{abstract}
\vspace{0.5cm}

\noindent \textbf{Abstract:} \revision{We consider a two-dimensional scalar field theory that modifies the standard $\phi^4$ model by introducing a smooth breaking of translational invariance through a hyperbolic generalizing function. This function explicitly breaks the $\mathds{Z}_2$ symmetry; however, it also introduces a mechanism capable of generating new energy minima (vacua) and of localizing or delocalizing field fluctuations around these vacua. Thus, this mechanism enables the continuous transformation of the kink/antikink-like configurations into compacted asymmetric double-kink/antikink structures. Accordingly, this transformation gives rise to new classes of configurations resembling asymmetric non-topological solitons, characterized by a soft breaking of the $\mathbb{Z}_2$ symmetry. These findings are particularly compelling, as they are consistent with results describing Su-Schrieffer-Heeger (SSH) domain walls in dimerized polymer systems.}

%\noindent \textbf{Keywords:} Modified $\phi^4$ theory. Extended classical solutions. Translational invariance.
\end{abstract}

\maketitle

\thispagestyle{empty}

\newpage 

\section{Introduction}

It has become evident that the study of domain walls, and consequently of kinks, has advanced significantly in both theoretical \cite{Abraham,Belendryasova,Bazeia,Andrade} and experimental \cite{KLi,Luv} physics. This field of research has established itself as a bridge between classical and quantum theories \cite{Evslin,Albayrak,Vachaspati}. These encompassing contexts range from condensed matter systems \cite{Braun,Byczuk} to gravitational \cite{CAlmeida,FBrito,FLima} theories. Initially, these structures emerge from investigations of the $\phi^4$ model \cite{Finkelstein}. In pursuit of this purpose, Finkelstein analyzes classical solitonic solutions, specifically kinks, which arise from the mechanism of spontaneous symmetry breaking in quartic $\phi^4$-like potentials \cite{Finkelstein}. In this framework, the field solutions connect different minima of the potential, thereby characterizing distinct classes of topological defects. In his studies, Finkelstein focuses on analyzing the conditions under which such defects, namely kinks, can exist, as well as on investigating their properties. He even proposes scenarios in which these topological defects exhibit characteristics analogous to those of fermionic particles \cite{Finkelstein}.

Generally speaking, kinks represent the simplest class of stable solitonic solutions\footnote{Kinks are not genuine solitons, as they undergo deformation when interacting with other kinks and/or antikinks \cite{Rajaraman}.} describing the fundamental examples of topological excitations \cite{Rajaraman,Manton}. Over time, this concept has been extended to domain walls in systems with broken symmetries, playing a crucial role in understanding phase transitions, defects in liquid crystals, and pattern formation in ferroelectric materials \cite{Handschy1,Handschy2,Clark1}. In high-energy physics and cosmology, e.g., domain walls are predicted as remnants of phase transitions in the early universe \cite{Kibble} and are relevant to the development of theories and models in these fields \cite{Vilenkin1,Vilenkin2,Vilenkin3,Ma}. This progression reflects not only a deepening of theoretical understanding but also a growing experimental \cite{Yamaguchi,Heeger,Toth} and computational \cite{Dorey,Han,AGomes} interest in kink-like topological structures as fundamental elements of nature.

The proposed $\phi^4$ theory makes the $\mathds{Z}_2$ symmetry an persistent feature from this model \cite{Coleman1,Coleman2,Goldstone,Dolan} once the action, in its simplest form, takes the profile
\begin{align}\label{SSS}
    S=\int\,d^2x\, \left[\frac{1}{2}\partial_\mu\phi\,\partial^\mu\phi-\frac{\lambda}{2}(\nu^2-\phi^2)^2\right],
\end{align}
which makes this discrete symmetry\footnote{One defines this symmetry for invariance under the discrete transformation $\phi\to-\phi$ \cite{Vachaspati,Manton,Rajaraman}.} a fundamental element in the formulation of two-dimensional scalar field theory is its profound role in shaping the model's vacuum structure. Moreover, when the potential of the theory, such as in the $\phi^4$ model, exhibits spontaneous symmetry breaking, multiple degenerate energy minima (vacua) emerge, as in $\phi = \pm\nu$. This feature enables the existence of classical solutions known as kinks (or solitons), which interpolate between different vacua supported by the model. For instance, in the theory described by Eq. \eqref{SSS}, we have $\phi_{-\infty} \to \mp\nu$ and $\phi_{+\infty} \to \pm\nu$, representing stable states due to the presence of an energy barrier separating the vacua \cite{Lima1,Lima2,Lima3}. In this conjecture, kinks are not only relevant within field theory itself but also appear in broader areas of physics, including condensed matter physics \cite{GLu,Buij,Rodrigues}, cosmology \cite{Skenderis}, and superstring theory \cite{Bergshoeff,MEto}. The study of such field configurations is essential, as the topology of the vacuum space, comprising two distinct points at spatial infinity, ensures the existence of topologically distinct classes of solutions. This topological protection prevents kink-like domain walls from continuously decaying into a trivial vacuum \cite{Vachaspati}.

Meanwhile, the class of genuine kink solutions exhibits translational invariance \cite{Vachaspati,Manton,Rajaraman}. This property means that such solutions remain physically equivalent under spatial translations \cite{Rajaraman}. In two-dimensional scalar field models with $\phi^4$ interaction, the kinks are solitonic solutions that interpolate between two distinct minima of the potential. Since the action is invariant under spatial translations. Thus, shifting the kink’s position does not affect the total energy. Consequently, the core position of the kink is a free parameter concerning a zero mode in perturbative analysis, reflecting the translational symmetry. This invariance is essential for semiclassical quantization and for understanding kink dynamics, particularly in collision processes and interactions with other field excitations \cite{Vachaspati}.

However, it is essential to highlight that the breaking of translational invariance in deformed kink-like configurations relates to the loss of the $\mathds{Z}_2$ symmetry\footnote{In the scalar field theory, the violation of the discrete $\mathds{Z}_2$ symmetry leads us to significant consequences for the framework of topological field solutions and the system’s translational symmetry. Especially, the breaking of this symmetry may induce asymmetries in the potentials, and thus in kink/antikink-like solutions, resulting in configurations that are either non-centered or not equivalent under spatial translations. Thus, the system may lose invariance under continuous translations, meaning that global translational symmetry is compromised. That also leads to modified zero modes, which allow the emergence of new dynamical properties in the solutions \cite{Rao,Watanabe}. These implications become especially relevant in effective models of materials and low-dimensional systems, where $\mathds{Z}_2$ symmetry breaking can be interpreted as a signature of external interactions or geometric deformations of space \cite{Ammon}.}, i.e., the violation of discrete symmetry \cite{Coleman1}. In a system exhibiting translational invariance, the solutions to the equations of motion do not depend on the absolute position but only on the relations between the boundary points (vacua) \cite{Coleman1,Coleman2}. Nonetheless, when a kink forms\footnote{Kinks are topologically nontrivial solitonic solutions that connect distinct vacua \cite{Finkelstein}.}, spontaneous symmetry breaking occurs. From this perspective, we characterized the kink as a spatially localized structure with a well-defined core, which implies that its translation physically alters the system, even though the total energy of the solution remains unchanged. That leads to the emergence of a zero mode associated with the translation of the kink, reflecting the continuous degeneracy of the solution \cite{Vachaspati}. Such a phenomenon has significant implications, including the presence of low-energy collective modes and the necessity for careful treatment in the quantization around the solitonic solution.

In the development of this work, our main purposes are to investigate new classes of solutions that may arise from a modified $\phi^4$ theory [Eq. \eqref{SSS}], i.e., a noncanonical version of the $\phi^4$ theory. In particular, we examine whether double kink/antikink-like solutions can emerge within a two-dimensional flat spacetime. Furthermore, we seek to understand how the deformation from kinks (antikinks) to double-kinks (double-antikinks) affects the translational invariance. Finally, we explore whether there are phenomenological results consistent with these theoretical field profiles. The goal of this work is precisely to address these questions throughout its formulation.

We outline this article into five distinct sections. Sect. \ref{sec1} presents the formulation of the two-dimensional noncanonical scalar theory. The equation of motion for an arbitrary interaction is derived and shown to admit a reduction in order, allowing the formulation of the Euler-Lagrange equation as a first-order differential equation. Naturally, this equation corresponds to the BPS equation of the noncanonical scalar field theory. Furthermore, we obtain the analytical expressions for the translational and vibrational modes for arbitrary interactions. Besides, we show the mapping of the noncanonical model for standard theory. In Sect. \ref{sec2}, we adopt the approach developed in Sect. \ref{sec1} and investigate the deformation mechanism of kink/antikink structures into asymmetric double-kink (double-antikink) configurations. Besides, one performs the inspection of the stability of the solutions and their implications for translational symmetry. We expose the phenomenological consistency of the model in Sect. \ref{sec3}. Additionally, one announces our findings in Sect. \ref{sec4}.

\vspace{-0.5cm}
\section{Framework of the Two-Dimensional Scalar Field Theory}\label{sec1}

Let us begin our work by considering a two-dimensional action of a non-canonical scalar field theory\footnote{A theory is considered non-canonical whenever its kinetic term deviates from the standard form $\frac{1}{2}\,\partial_\mu\phi\,\partial^\mu\phi$.} governed by a field $\phi(x,t)$ in flat spacetime\footnote{One adopts the metric signature of spacetime as $\eta_{\mu\nu} = \text{diag}(+,-)$.} \cite{Lima1,Lima2,Lima3}, viz.,
\begin{align}\label{Eq1}
    S=\int\, d^2x\,\left[\frac{f(\phi)}{2}\partial_\mu\phi\,\partial^\mu\phi-V(\phi)\right].
\end{align}
Naturally, $\phi(x,t)$ denotes a real scalar field with $t$ representing the temporal coordinate, $x$ the spatial coordinate, and $f(\phi)$ a positive-definite generalizing function. This function is responsible for breaking the $\mathbb{Z}_2$ symmetry. The term $V(\phi)$ corresponds to the potential, while $\frac{1}{2}\,\partial_\mu\phi\,\partial^\mu\phi$ represents the standard kinetic contribution. From a physical standpoint, $f(\phi)$ effectively modifies the potential $V(\phi)$, potentially leading to new phenomena such as the emergence of additional vacua, the displacement of existing vacuum positions, and the decay of false vacuum states\footnote{One can map the noncanonical model into a canonical theory. We will expose this mapping soon.}.

We now investigate the equation of motion corresponding to the theory displayed in Eq. \eqref{Eq1}. For this purpose, let us apply the principle of least action, i.e., $\delta S = 0$, which results in
\begin{align}\label{Eq2}
    f\partial_\mu\partial^\mu\phi+\frac{f_\phi}{2}\partial_\mu\phi\,\partial^\mu\phi+V_\phi=0,
\end{align}
where $f_\phi$ and $V_\phi$ denote, respectively, the partial derivatives of the functions $f(\phi)$ and $V(\phi)$ concerning the scalar field $\phi$. Mathematically, these notations correspond to $f_\phi=\frac{\partial f}{\partial \phi}$ and $V_\phi=\frac{\partial V}{\partial \phi}$.

Naturally, one can reformulate Eq. \eqref{Eq2} as
\begin{align}\label{Eq3}
    f\ddot{\phi}-f\phi''+\frac{f_\phi}{2}\dot{\phi}^2-\frac{f_\phi}{2}\phi'\,^2+V_\phi=0.
\end{align}
In this framework, the prime and dot notations denote derivatives concerning position and time, respectively. 

Hereafter, let us consider the static case ($\dot{\phi} = 0$). Thereby, the equation of motion \eqref{Eq3} reduces to 
\begin{align}
    \label{Eq4}
    f\phi''+\frac{f_\phi}{2}\phi'\,^2=V_\phi.
\end{align}
By multiplying Eq. \eqref{Eq4} by $\phi'$, one obtains
\begin{align}
    \label{Eq5}
    \frac{d}{dx}\left(\frac{f\phi'\,^2}{2}-V\right)=0,
\end{align}
which leads us to the first-order equation\footnote{In this case, we assume that the constant arising from Eq. \eqref{Eq5} is zero.}
\begin{align}
    \label{Eq6}
    \phi'=\pm\sqrt{\frac{2V}{f}}.
\end{align}
Eq. \eqref{Eq6} is the BPS equation\footnote{The term BPS refers to the Bogomol’nyi \cite{Bogomolnyi}, Prasad, and Sommerfield \cite{PrasadS}, who developed the formulation of classical field solutions with special properties. This approach describes stable solutions or low-energy configurations that preserve the symmetries while simultaneously reducing the order of the equations of motion.}. Unsurprisingly, since $V(\phi)$ is a function solely of $\phi$, one can express the potential $V(\phi)$ in terms of a new function that depends only on the scalar field $\phi$. We refer to this new function as $W(\phi)$ (or the superpotential). Accordingly, we conveniently write the potential as
\begin{align}
    \label{Eq7}
    V(\phi)=\frac{W_{\phi}^2}{2f},
\end{align}
which allows us to reduce Eq. \eqref{fig6} to
\begin{align}
    \label{Eq8}
    \phi'=\pm\frac{W_\phi}{f}.
\end{align}
Here, the notation $W_\phi$ is the derivative of the function $W(\phi)$ concerning scalar field $\phi$ \footnote{For further details on the superpotential (or auxiliary function) $W(\phi)$, see Ref. \cite{Adam0}.}. Meanwhile, the Eq. \eqref{Eq8} corresponds to the BPS equation \cite{Bogomolnyi,PrasadS}.

Within this framework, the energy of the real scalar field is
\begin{align}\label{Eq9} \nonumber
    \text{E} =&\int_{-\infty}^{\infty}\,dx\, \left[-\frac{f(\phi)}{2}\partial_\mu\phi\,\partial^\mu\phi+V(\phi) \right]\\
    \text{E} =&\int_{-\infty}^{\infty}\,dx\, \left[-\frac{f(\phi)}{2}\dot{\phi}\,^2+\frac{f(\phi)}{2}\phi'\,^2+V(\phi)\right].
\end{align}
Therefore, in the static case, adopting Eqs. \eqref{Eq7} and \eqref{Eq8}, the BPS energy reduces to
\begin{align}
    \label{Eq10} 
    \text{E}_{\text{BPS}}= \int_{-\infty}^{\infty}\, \left(\frac{W_{\phi}^{2}}{f}\right)\, dx=\pm\int_{-\infty}^{\infty}\, W_\phi\, \phi'\,dx=\pm\int_{\phi_{-\infty}}^{\phi_{\infty}}\,W_{\bar{\phi}}\, d\bar{\phi},
\end{align}
therefore,
\begin{align}
    \label{Eq11}
    \text{E}_{\text{BPS}}= &\pm[W(\phi_{+\infty})-W(\phi_{-\infty})].
\end{align}
Here, it is adopted the notation $W_{\phi_{\pm\infty}}=W[\phi(x\to\pm\infty)]$.

The analysis of Eq. \eqref{Eq10} allows us to conclude that the BPS energy density takes the form
\begin{align}
    \label{Eq12}
    \mathcal{E}_{\text{BPS}}=\frac{W_{\phi}^{2}}{f}.
\end{align}

Furthermore, it is essential to highlight that the constraints
\begin{align}
    \lim_{\vert x\vert\to\infty}\sqrt{f(\phi)}\phi'=0 \hspace{0.5cm} \text{and} \hspace{0.5cm} \lim_{\vert x\vert\to \infty}V(\phi)=0
\end{align}
ensure that the energy [Eq. \eqref{Eq9}] remains finite.

\revision{Let us highlight that the real scalar field satisfies the topological boundary conditions, viz., 
\begin{align}
    \phi_{-\infty}=\mp\nu \hspace{1cm} \text{and} \hspace{1cm} \phi_{+\infty}=\pm\nu.
\end{align}
In this framework, the upper sign corresponds to the boundary conditions of kink-like solutions, whereas the lower sign characterizes antikink-like solutions\footnote{\revision{Mathematically, $\phi_{\mp\infty}$ corresponds, respectively, to $\phi_{-\infty}\equiv\phi(x\to -\infty)$ and $\phi_{+\infty}\equiv\phi(x\to +\infty)$.}}.}

\subsection{The consistency of the BPS property}\label{sec1a}

To verify the consistency of Eqs. \eqref{Eq8} and \eqref{Eq11}, let us recall that the energy density is defined as the integral over the entire space of the $T_{00}$ component of the energy-momentum tensor\footnote{The energy-momentum tensor is defined as $T^{\mu}\,_{\nu}=\frac{\partial\mathcal{L}}{\partial(\partial_\mu\phi)}\partial_\nu\phi-\delta^{\mu}\,_{\nu}\,\mathcal{L}$. In the two-dimensional case, the component $T^{0}\,_{0}$ boil down to $T^{0}\,_{0} = -\mathcal{L}$. For further details, see Ref. \cite{Rubakov}.}. Accordingly, the energy of the scalar field is
\begin{align}\label{Eq13}
    \text{E}=& \int_{-\infty}^{+\infty}\, dx\, \left[\frac{f}{2}\dot{\phi}\,^2+\frac{f}{2}\phi'\,^2+V\right].
\end{align}
Therefore, to analyze whether the model admits the BPS property, we introduce an auxiliary function known as the superpotential, i.e., $W \equiv W(\phi)$, which plays a fundamental role once directly related to the potential and total energy, providing a convenient and advantageous approach \cite{Adam0}. Accordingly, by reformulating the expression \eqref{Eq13}, one obtains
\begin{align}
    \label{Eq14}
    \text{E}= \int_{-\infty}^{\infty}\, dx\, \left[\frac{f}{2}\dot{\phi}\,^2+\frac{1}{2f}\left(f\phi'\mp W_\phi\right)^2+V-\frac{W_{\phi}^2}{2f}\pm\frac{d W}{dx} \right].
\end{align}
As a result, for the model to have BPS property, the potential must be 
\begin{align}
    \label{Eq15}
    V=\frac{W_{\phi}^2}{2f},
\end{align}
which allows us to simplify the energy to
\begin{align}
    \label{Eq16}
    \text{E}= \int_{-\infty}^{\infty}\, dx\, \left[\frac{f}{2}\dot{\phi}\,^2+\frac{1}{2f}\left(f\phi'\mp W_\phi\right)^2\pm\frac{d W}{dx} \right].
\end{align}

Naturally, we can also express the total energy as 
\begin{align}
    \label{Eq17}
    \text{E}= \text{E}_{\text{BPS}}+\int_{-\infty}^{\infty}\, dx\,\left[\frac{f}{2}\dot{\phi}^2+\frac{1}{2f}\left(f\phi'\mp W_\phi\right)^2\right],
\end{align}
 where $\text{E}_{\text{BPS}}$ is the BPS bound, viz.,
\begin{align}\label{Eq18}
    \text{E}_{\text{BPS}}=\pm\,\int_{-\infty}^{\infty}\, \frac{d W}{d x}\, dx=\pm[W(\phi_{+\infty})-W(\phi_{-\infty})]>0.
\end{align}
Unsurprisingly, one notes that $\mathrm{E}\geq \mathrm{E}_{\mathrm{BPS}}$. Thus, the energy bound, i.e., $\mathrm{E} = \mathrm{E}_{\mathrm{BPS}}$, the self-dual equation adopt the form
\begin{align}
    \label{Eq19}
    \dot{\phi}=0 \hspace{0.5cm} \text{and} \hspace{0.5cm} \phi'=\pm\frac{W_\phi}{f}.
\end{align}
These are the static self-dual differential equations. Therefore, one notes the consistency of Eqs. \eqref{Eq7} and \eqref{Eq8} with the BPS theory [Eqs. \eqref{Eq15} and \eqref{Eq19}].

\subsection{On the linear stability}\label{sec1b}

Let us now analyze the stability of the field configurations that satisfy the equation of motion \eqref{Eq8}. Towards this objective, let us consider that $\phi_s(x)$ is a solution that satisfies Eq. \eqref{Eq8}. Taking this into account, we introduce small perturbations of the form $\eta(x)\cos(\omega t)$ around the solution $\phi_s(x)$. That immediately yields
\begin{align}\label{Eq20}
    \phi(x,t)=\phi_s(x)+\eta(x)\cos(\omega t) \hspace{0.5cm} \text{with} \hspace{0.5cm} \vert\vert\eta(x)\vert\vert\ll\vert\vert\phi_s(x)\vert\vert.
\end{align}
\revision{Here, the field $\phi(x,t)$ evolves according to the equation of motion \eqref{Eq3}. Therefore, by substituting the expansion \eqref{Eq20} into \eqref{Eq3} and considering only linear perturbations, one arrives at
\begin{align}
    \label{Eq21}
    -\frac{d^2\eta}{dx^2}-\frac{1}{f}\frac{\partial f}{\partial \phi}\bigg\vert_{\phi=\phi_s}\left(\frac{d\phi_s}{dx}\right)\frac{d\eta}{dx}+U_0(x)\eta=\omega^2\eta,
\end{align}
where $\omega\geq 0$ and $U_0(x)$ is
\begin{align}
    \label{Eq22}
    U_0(x)=\frac{1}{f}\frac{\partial^2V}{\partial\phi^2}\bigg\vert_{\phi=\phi_s}-\frac{1}{f}\frac{\partial f}{\partial\phi}\bigg\vert_{\phi=\phi_s}\frac{d^2\phi_s}{dx^2}-\frac{1}{2f}\frac{\partial^2f}{\partial\phi^2}\left(\frac{d\phi_s}{dx}\right)^2.
\end{align}
Therefore, one infers that the stability of the solutions $\phi_s(x)$ will depend on the eigenvalues being real \footnote{\revision{One highlights that the stability of the solutions depends on the positivity of the spectrum of linear fluctuations around the classical field configurations [i.e., Eq. \eqref{Eq20}]. Accordingly, for the configurations to be stable, it is required that $\omega_n^2 \geq 0$ for all $n$. In other words, there must be no modes with $\omega_n < 0$, as these would correspond to fluctuations that grow exponentially in time \cite{Manton,Vachaspati,Rajaraman}.}}.}

\subsubsection{Description of the zero mode (translational mode)}\label{sec1b1}

Since the field configuration $\phi_s(x)$ satisfies the static Euler–Lagrange equation \eqref{Eq4}, to determine the eigenfunction $\eta_0(x)$ associated with the zero mode ($\omega = 0$), we differentiate Eq. \eqref{Eq4} concerning the spatial coordinate. That yields
\begin{align}
    \label{Eq23} \nonumber
    -f\frac{d^2\phi_s'}{dx^2}-\frac{\partial f}{\partial\phi}\bigg\vert_{\phi=\phi_s}\left(\frac{d^2\phi_s'}{dx^2}\right)\phi_s'-\frac{1}{2}\frac{\partial^2 f}{\partial\phi^2}\bigg\vert_{\phi=\phi_s}\left(\frac{d\phi_s}{dx}\right)^2\phi_s'-\frac{\partial f}{\partial \phi}\bigg\vert_{\phi=\phi_s}\left(\frac{d\phi_s'}{dx}\right)\phi_s'+&\\
    +\frac{\partial^2V}{\partial \phi^2}\bigg\vert_{\phi=\phi_s}\phi_s'=0&,
\end{align}
where $\phi_s'$ denotes the derivative of $\phi_s$ concerning the position variable.

Rearranging the above expression, one notes that Eq. \eqref{Eq23} reduces to
\begin{align}
    \label{Eq24}
    -\frac{d^2\phi_s'}{dx^2}-\frac{1}{f}\frac{\partial f}{\partial\phi}\bigg\vert_{\phi=\phi_s}\left(\frac{d\phi_s}{dx}\right)\frac{d\phi_s'}{dx}+U_0(x)\phi_s'=0,
\end{align}
where
\begin{align}
    U_0(x)=\frac{1}{f}\frac{\partial^2V}{\partial\phi^2}\bigg\vert_{\phi=\phi_s}-\frac{1}{f}\frac{\partial f}{\partial\phi}\bigg\vert_{\phi=\phi_s}\left(\frac{d^2\phi_s}{dx^2}\right)-\frac{1}{2f}\frac{\partial^2f}{\partial\phi^2}\bigg\vert_{\phi=\phi_s}\left(\frac{d\phi_s}{dx}\right)^2.
\end{align}

By comparing the identity \eqref{Eq24} with the perturbed equation \eqref{Eq21}, it follows that the zero mode $\eta_0(x)$, i.e., the eigenfunction associated with the eigenvalue $\omega=0$ is 
\begin{align}
    \label{Eq25}
    \eta_{0}(x)=\frac{d\phi_s}{dx}.
\end{align}

\subsubsection{The case of non-zero modes (vibrational modes)}\label{sec1b2}

To investigate the vibrational modes, i.e., those with non-zero eigenvalues ($\omega\neq 0$), let us consider Eq. \eqref{Eq21} and apply the change of variable
\begin{align}
    \label{Eq26}
    \eta(x)=f^{-1/2}(\phi_s)\Psi(x).
\end{align}
Naturally, the change of variable [Eq. \eqref{Eq26}] leads to a Schr\"{o}dinger-like equation, viz., 
\begin{align}
    \label{Eq27}
    \hat{H}\Psi(x)=\omega^2 \Psi(x),
\end{align}
such that $\omega^2$ is the eigenvalue and $\hat{H}$ is
\begin{align}
    \label{Eq28}
    \hat{H}=-\frac{d^2}{dx^2}+U_{\text{eff}}(x),
\end{align}
where the effective stability potential is
\begin{align}
    \label{Eq29}
    U_{\text{eff}}(x)=\frac{1}{f}\frac{\partial^2V}{\partial\phi^2}\bigg\vert_{\phi=\phi_s}-\frac{1}{2f}\frac{\partial f}{\partial \phi}\bigg\vert_{\phi=\phi_s}\frac{d^2\phi_s}{dx^2}-\frac{1}{4f^2}\left(\frac{\partial f}{\partial\phi}\right)^2\bigg\vert_{\phi=\phi_s}\left(\frac{d\phi_s}{dx}\right)^2.
\end{align}

\subsection{From the non-canonical theory to the canonical theory}\label{sec1c}

Naturally, one can map the non-canonical theory [Eq. \eqref{Eq1}] into a canonical theory (or na\"{i}ve), i.e., a theory with a kinetic term in the standard form. To demonstrate this possibility, let us consider the action
\begin{align}
    \label{Eq30}
    S=\int\, d^2x\, \left[\frac{f(\phi)}{2}\partial_\mu\phi\,\partial^\mu\phi-V(\phi)\right].
\end{align}
Here, the $f(\phi)$ is a positive-define generalizing function. Furthermore, one must choose a $f(\phi)$-profile to ensure the absence of ghosts, i.e., field configurations with negative energy (and energy density). Taking that into account, let us perform the mapping of the scalar field through the transformation \cite{Losano}, viz.,
\begin{align}
    \label{Eq31}
    \bar{\Phi}=\int^{\phi}\, d\tilde{\phi}\, \sqrt{f(\tilde{\phi})}\equiv h(\phi).
\end{align}

Thus, the transformation \eqref{Eq31} indicates that
\begin{align}\label{Eq32}
    h_\phi=\sqrt{f(\phi)},
\end{align}
which allows us to conclude that
\begin{align}
    \label{Eq33}
    \partial_\mu\bar{\Phi}\equiv h_\phi\partial_\mu\phi\to \partial_\mu\bar{\Phi}\,\partial^\mu\bar{\Phi}\equiv h_{\phi}^{2}\,\partial_\mu\phi\,\partial^\mu\phi\,\to \partial_{\mu}\bar{\Phi} \partial^{\mu}\bar{\Phi}\equiv f(\phi)\partial_\mu\phi\,\partial^\mu\phi.
\end{align}
Therefore, the effective canonical action corresponding to the action in Eq. \eqref{Eq30} is
\begin{align}
    \label{Eq34}
    S_{\text{eff}}=\int\, d^2x\, \left[\frac{1}{2}\partial_\mu\bar{\Phi}\,\partial^\mu\bar{\Phi}-V(h^{-1}(\bar{\Phi}))\right].
\end{align}
Note that, adopting the field transformation $\bar{\Phi}\to\int^{\phi}\,d\tilde{\phi}\,\sqrt{f(\tilde{\phi})}$ [Eq. \eqref{Eq31}], the originally non-canonical kinetic term becomes canonical. However, the potential, initially defined as $V(\phi)$, undergoes a deformation and is rewritten as $V(h^{-1}(\phi))$. Therefore, the two theories are equivalent, as both models describe physically equivalent solutions.

Indeed, there are two main approaches for implementing deformation mechanisms in field configurations in the literature. The first involves modifying the potential \cite{Malbouisson,Guilarte}, while the second acts directly on the field profiles through generalizations applied to the real scalar field or the electromagnetic sector \cite{Lima6,FCLima0}. Among these, the generalization-based approach has gained greater prominence \cite{Lima1,Lima2,Lima3,FCLima}. That is because the generalizing function $f(\phi)$ adjusts the contribution of the kinetic term and relates to the potential in regimes of saturated energy, thereby modifying the vacuum structure of the theory and enabling the deformation of the field profiles.

\revision{Furthermore,  it is worth emphasizing that the non-canonical theory obtained through the generalizing function $f(\phi)$, coupled to the kinetic term, is not merely a mathematical abstraction but reflects the physical behavior of real systems. In cosmology, for instance, this framework arises in models of inflation and dark energy. In this framework, the generalizing function defines an effective metric in the internal field space, rendering it curved. That implies that the field does not propagate in a free medium, but in an environment where the very geometry of field space induces effective interactions \cite{Picon,Garriga}. Meanwhile, in condensed matter physics, the function $f(\phi)$ accounts for anisotropic media, critical transitions, and collective excitations \cite{Chaikin,Altland}. In this setting, the propagation of excitations, such as spin waves, phonons, or collective modes, depends on the background density or on the order parameter, and is suitably described by a multiplicative factor in the kinetic term (i.e., the generalizing function) \cite{Chaikin,Altland}. Moreover, this generalizing function provides a mechanism for tuning phase transitions through the effective stiffness.}

\section{Hyperbolically deforming domain walls}\label{sec2}

In this section, we present the numerical method used to obtain new results for a modified $\phi^4$ theory, in which the generalizing function $f(\phi)$ assumes a hyperbolic profile.

\subsection{Computational approach}\label{sec2a}

To achieve our purpose, we will estimate the solutions of the BPS equations \eqref{Eq8} and \eqref{Eq19} using the numerical interpolation method. Particularly, this approach is essential once it provides the numerical solutions for the differential equations \eqref{Eq8} and \eqref{Eq19}. In general, this method involves discretizing the domain of the function in a differential equation of the form
\begin{align}
    \label{Eq35}
    \frac{df(x)}{dx}=g(f(x)).
\end{align}
This discretization assumes that the domain of Eq. \eqref{Eq35} consists of a well-defined and discrete range for the position variable $x$, allowing $x$ to be treated as discrete and the intervals to be clearly defined. Naturally, we discretize the position coordinate in Eq. \eqref{Eq35} [and, by extension, in Eqs. \eqref{Eq8} and \eqref{Eq19}] by subdividing the range into ten thousand discrete points. Thus, the solution to the differential equation is investigated within the interval $[-10, 10]$, using the discretization $x_1,\, x_2,\,\dots ,\, x_{10000}$, where $x_{1}=-10$ and $x_{10000}=10$.

By performing the discretization of the position variable and employing the Euler method \cite{Butcher,Sauer}, viz.,
\begin{align}
    \label{Eq36}
    f_{n+1}=f_n+m\cdot \lambda(x_n,f_n).
\end{align}
Thus, we estimate the solution $f(x)$ point by point. In this framework, $m=10^{-10}$ represents the step size, and $\lambda(x_n,\, f_n)$ is the derivative of the function at the point $(x_n,\, f_n)$.

In summary, one estimates the values between discrete points through polynomial interpolation. It is essential to highlight that interpolation is an approach used to construct a continuous function that passes exactly through the numerically computed points. In the overview, we employed low-degree polynomial functions to ensure smooth transitions between discrete points and to guarantee the existence of derivatives. For further details, see Refs. \cite{Scarborough,Burden}.

 \subsection{The hyperbolically modified $\phi^4$ theory}\label{sec2b}

\subsubsection{The system setup}\label{sec2b1}

To pursue our line of inquiry, it is necessary to select a profile for the superpotential $W(\phi)$ and a generalizing function $f(\phi)$. Naturally, it is convenient to adopt a superpotential that preserves the mechanism of spontaneous symmetry breaking while maintaining the $\mathds{Z}_2$ symmetry manifest when $f(\phi) \to 1$. Accordingly, we select a superpotential that describes the $\phi^4$ theory, viz.,
\begin{align}
    \label{Eq37}
    W(\phi)=\sqrt{\lambda}\left(\nu^2\phi-\frac{\phi^3}{3}\right),
\end{align}
where $\lambda$ is a coupling parameter and $\nu$ the Vacuum Expectation Value (VEV).

The choice of the $\phi^4$ model is particularly convenient, as it constitutes the most fundamental and well-understood model among scalar field theories that exhibit nontrivial topological two-dimensional solutions. Moreover, the $\phi^4$ theory holds significant relevance due to its relation to statistical models that describe phase transitions and critical phenomena, such as the Ising model \cite{Ising, Mayo, Karma}. Furthermore, one finds quite studies about the $\phi^4$ theory in the literature. For instance, we find the $\phi^4$ model in studies on new classes of topological solutions \cite{Lima1,Lima2,Lima3,Bazeia1,Bazeia2,Simas1} and the analysis of scattering processes involving topological structures \cite{Simas2}, among others.

 In addition to considering the superpotential \eqref{Eq37}, let us select a profile for the generalizing function $f(\phi)$. In this case, we choose a function that explicitly breaks the $\mathds{Z}_2$ symmetry\footnote{\revision{Naturally, we adopt this choice because it induces asymmetries in the vacuum state of the theory, while also allowing for the emergence of new vacua \cite{Lima1,Lima2,Lima3}. From a physical viewpoint, this feature is particularly relevant, once the omnipresence of the $\mathbb{Z}_2$ symmetry breaking plays a central role in condensed matter physics, being directly concerned with the appearance of ordered phases and the occurrence of critical transitions \cite{Stanley,Lines}. In general terms, the $\mathbb{Z}_2$ symmetry corresponds to a sign inversion operation in an order parameter, such that its violation characterizes the spontaneous selection among degenerate states. This macroscopic choice gives rise to nontrivial collective properties uninferable from the microscopic dynamics; instead, they emerge from the collective organization of the system. For instance, in one-dimensional systems and conjugated polymers, such as in the Su-Schrieffer-Heeger (SSH) model, the $\mathbb{Z}_2$ symmetry governs the selection between two equivalent lattice dimerization patterns \cite{SSH1}. It's breaking thus leads to the opening of an electronic gap and the consequent formation of new topological edge states, as well as the emergence of multiple domain walls.}}, namely,
\begin{align}
    \label{Eq38}
    f(\phi)=[\sinh(p\,\phi)+\cosh(p\,\phi)]^{-2}\sinh(q\,\phi^2)^{-\frac{1}{2z+1}}.
\end{align}

Here, the canonical dimensions are given by $[p]\equiv [\phi^{-1}]$ and $[q]\equiv [\phi^{-2}]$, with $z\in \mathds{Z}^{+}$ and dimensionless. Furthermore, the parameter $p$ is responsible for the asymmetry introduced via the breaking of the $\mathds{Z}_2$ symmetry\footnote{One can find further details regarding the breaking of the $\mathds{Z}_2$ symmetry and the loss of translational invariance in Refs. \cite{Rao,Watanabe,Ammon}.}. Meanwhile, $q$ adjusts the canonical dimension of the generalizing function without producing any direct physical effect. Finally, varying the parameter $z$ shifts the positions of the vacua state, directly influencing the deformation of the topological structures.

Naturally, the hyperbolic generalizing function proves to be both interesting and appropriate once Eq. \eqref{Eq15} [and, consequently, Eq. \eqref{Eq34}] indicates that, upon adopting the superpotential from Eq. \eqref{Eq37}, one obtains a hyperbolically modified $\phi^4$ theory. That modified $\phi^4$ theory becomes relevant due to its connection with the description of ferroelectric and ferromagnetic materials \cite{Lima2,Simas3}. Additionally, hyperbolic theories effectively describe the fragility and increased rigidity of crystalline lattices in several materials, for instance, ferroelectric materials \cite{Lima2,Simas3}. Beyond these applications, hyperbolic models also represent extensions of the Calogero model \cite{Calogero}, allowing for the formation of multiple confined solitons \cite{Gon}. Besides, this type of interaction arises in several physical branches, including supersymmetric theories \cite{Fedoruk}, compact objects \cite{QWen}, and inflationary models \cite{Agarwal}.

\begin{figure}[!ht]
  \centering
  \subfigure[The case without asymmetry ($p=0$).]{\includegraphics[height=5.5cm,width=5.4cm]{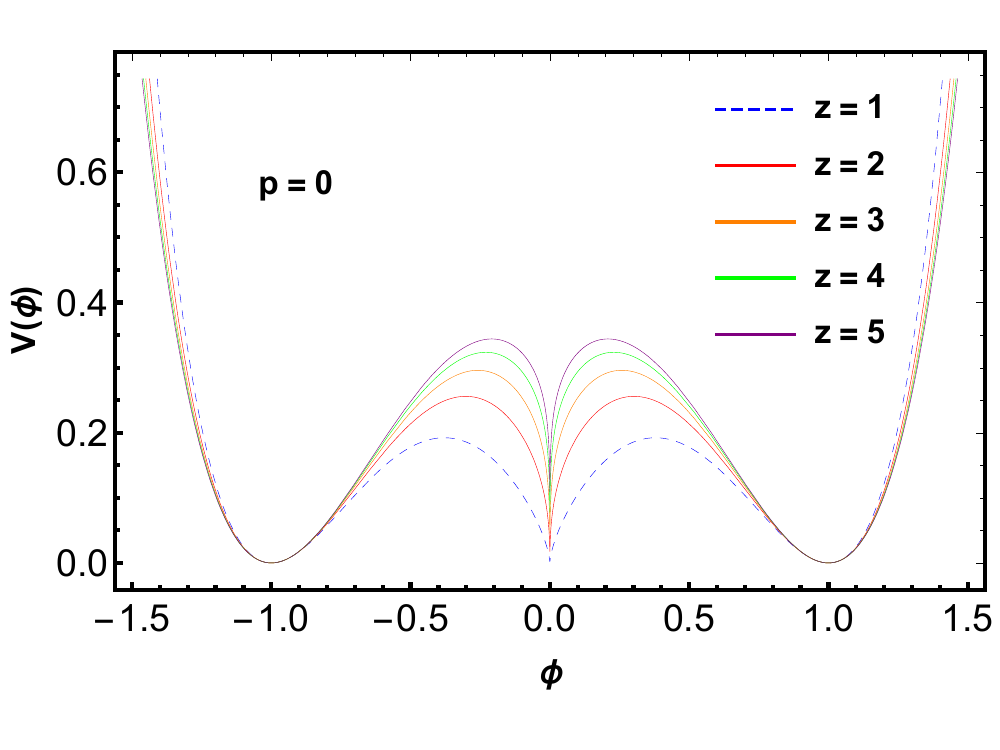}}
  \subfigure[The case with left-handed asymmetry ($p=1$).]{\includegraphics[height=5.5cm,width=5.4cm]{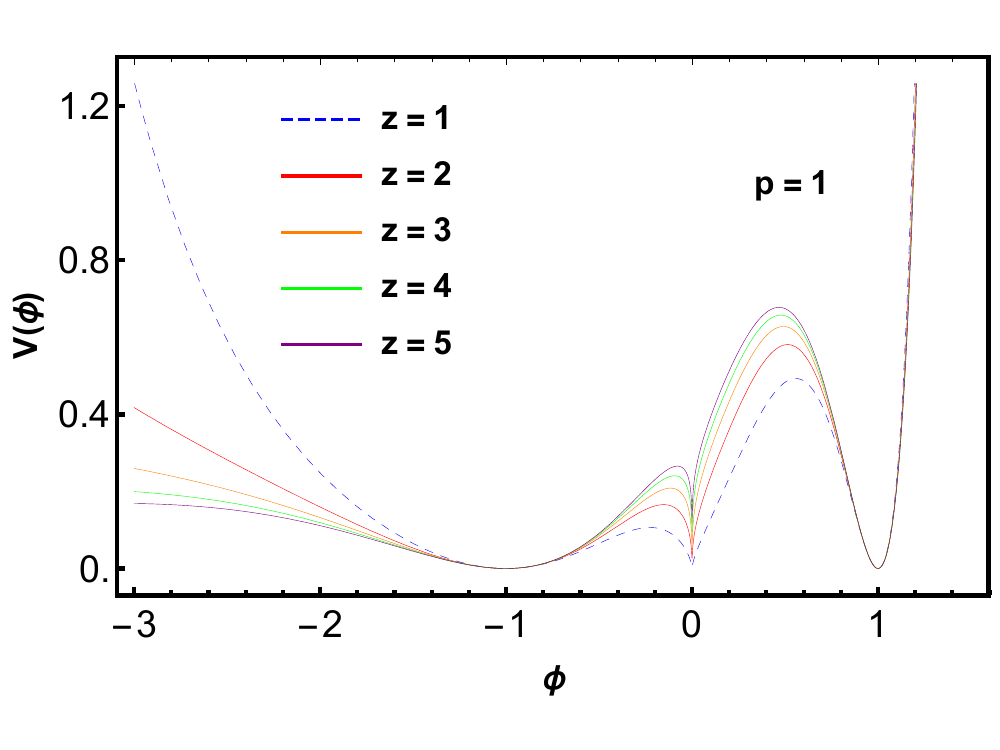}}
  \subfigure[The case with right-handed asymmetry ($p=-1$).]{\includegraphics[height=5.5cm,width=5.4cm]{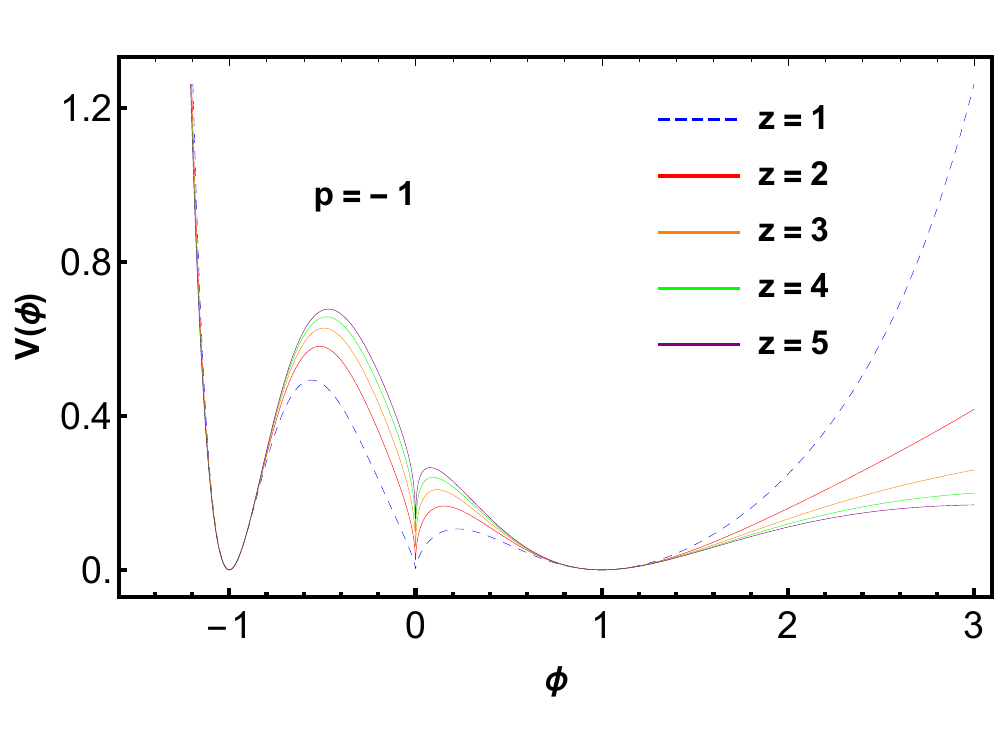}}\
  \caption{Potential  $V(\phi)$ vs. $\phi$ varying the parameter $z$ while keeping $p$ constant. In all case, we adopt $\lambda=\nu=q=1$.}
  \label{fig1}
\end{figure}

\begin{figure}[!ht]
  \centering
  \subfigure[The case with right-handed asymmetry for $z=1$ and $p<0$.]{\includegraphics[height=5.5cm,width=5.4cm]{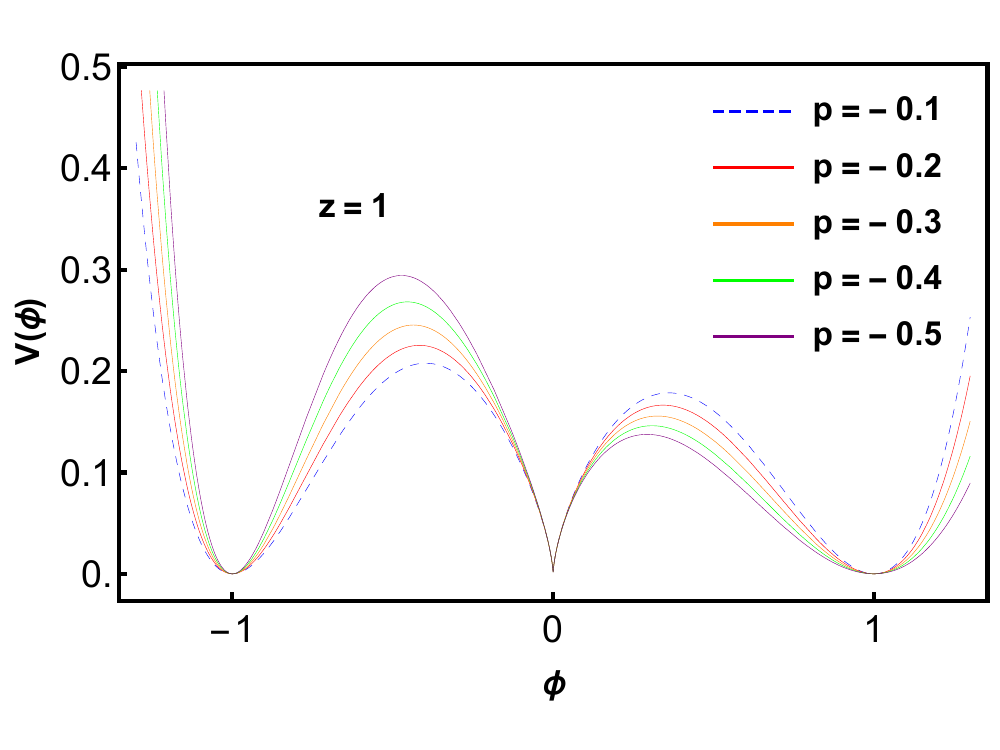}}\hspace{1cm}
   \subfigure[The case with left-handed asymmetry for $z=1$ and $p>0$.]{\includegraphics[height=5.5cm,width=5.4cm]{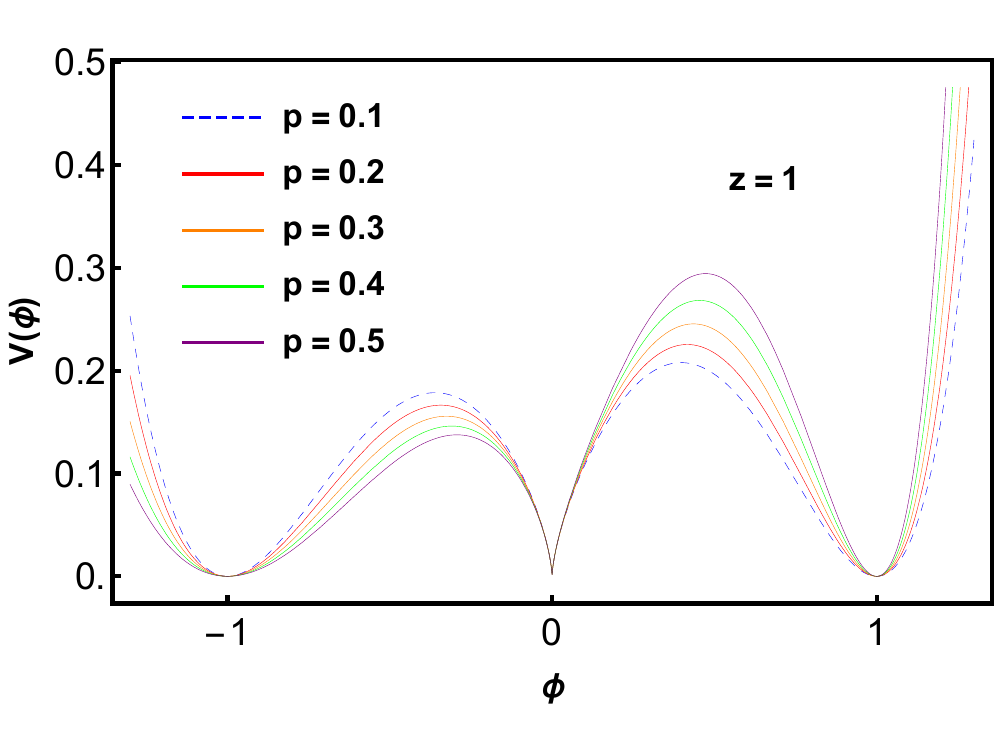}}\hfill
  \caption{Potential  $V(\phi)$ vs. $\phi$ varying the parameter $p$ while keeping $z$ constant. In all cases, we adopt $\lambda=\nu=q=1$.}
  \label{fig2}
\end{figure}

By adopting the superpotential [Eq. \eqref{Eq37}], the generalizing function [Eq. \eqref{Eq38}], and use the potential defined in Eq. \eqref{Eq7}, one obtains
\begin{align}
    \label{Eq39}
    V(\phi)=\frac{\lambda}{2}(\nu^2-\phi^2)^2[\sinh(p\,\phi)+\cosh(p\,\phi)]^2\sinh(q\,\phi^2)^{\frac{1}{2z+1}}.
\end{align}
Note that Eq. \eqref{Eq39} describes a hyperbolically modified $\phi^4$ potential. Naturally, the modified $\phi^4$ theory proves to be appropriate, as it provides a low-energy description for certain microscopic models. For instance, one justifies this approach by the fact that imperfections at the microscopic level often manifest as variations in the shape of the potential \cite{Lizunova,Abdullaev,Bilas,YZhou}. For more details on the deformations of the potential [Eq. \eqref{Eq39}], see Figs. \ref{fig1} and \ref{fig2}.

By considering Eq. \eqref{Eq8} with the potential defined in Eq. \eqref{Eq39}, one writes the equation of motion as
\begin{align}
    \label{Eq40}
    \phi'=\pm\sqrt{\lambda}(\nu^2-\phi^2)[\sinh(p\,\phi)+\cosh(p\,\phi)]^2\sinh(q\, \phi^2)^{\frac{1}{2z+1}}.
\end{align}
Here, $\phi'=\frac{d\phi}{dx}$. Moreover, the positive and negative signs correspond, respectively, to kink- and antikink-like solutions in the standard $\phi^4$ theory, i.e., when $f(\phi)\to 1$.

\subsubsection{Asymptotic behavior of the solutions}\label{sec2b2*}

Once the BPS equation \eqref{Eq40} is known, we can analyze the behavior of the solutions near the vacuum values to ensure the existence of a family of solutions. Towards this purpose, we examine the asymptotic behavior of the field in the limits $x \to -\infty$, $x \to 0$, and $x \to +\infty$. This analysis gives rise to the following asymptotic behaviors:
\begin{align}\label{An1}
    &\phi(x)\approx\mp\nu\pm\mathcal{C}_0\text{e}^{-M\,\vert x\vert}, \hspace{0.5cm} x\to-\infty,\\ \label{An2}
    &\phi(x)\approx \pm\left[\frac{(2z+3)\nu^2\sqrt{\lambda}\,x}{(2z+1)q^{\frac{1}{2z+1}}}\right]^{\frac{2z+1}{2z+3}}, \hspace{0.5cm} x\to 0,
\end{align}
and
\begin{align}\label{An3}
    \phi(x)\approx\pm\nu\mp\mathcal{C}_0\text{e}^{-M\,\vert x\vert}, \hspace{0.5cm} x\to+\infty, 
\end{align}
where $\mathcal{C}_0>0$, and the parameter $M$ is the BPS mass of the solitonic configurations, which in this scenario takes the form
\begin{align}\label{An4}
    \revision{M=2\nu\,\sqrt{\lambda}\,\text{e}^{2\, p\, \nu}\sinh(q\, \nu^2)^{-\frac{1}{2z+1}}.}
\end{align}

Based on the asymptotic behaviors [Eqs. \eqref{An1}, \eqref{An2}-\eqref{An3}], one notes that the profiles preserve the profile exponential decay of the naïve $\phi^4$ theory. However, the behavior at the neighborhood from the origin strongly suggests the presence of a deformation responsible for the emergence of a new class of double kink/antikink-like solutions. Naturally, the results indicate that the tails of the domain walls in the modified $\phi^4$ theory continue to decay exponentially. Thus, the central region of the solutions begins to exhibit features of compact-like profiles, with significant deformations around the origin. Therefore, we note that these behaviors are affected by the deformation ($z$) and asymmetry ($p$) parameters.

\subsubsection{The numerical solution}\label{sec2b2}

\begin{figure}[!ht]
  \centering
  \subfigure[Double-kink-like solutions.]{\includegraphics[height=5.5cm,width=5.4cm]{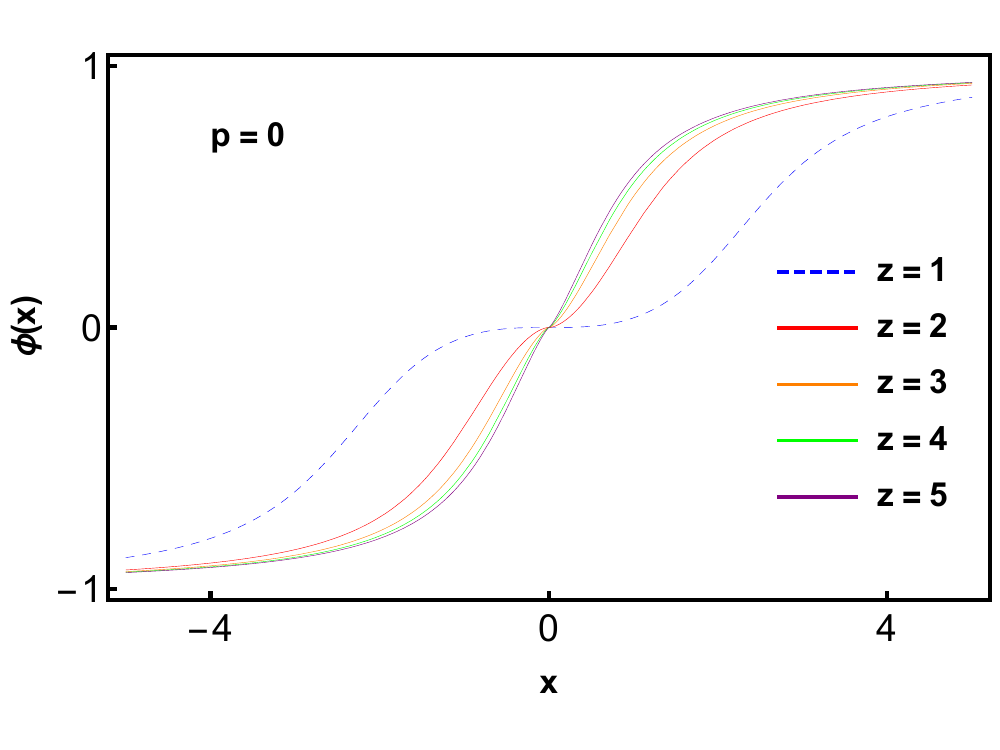}}\hspace{1cm}
   \subfigure[Double-antikink-like solutions.]{\includegraphics[height=5.5cm,width=5.4cm]{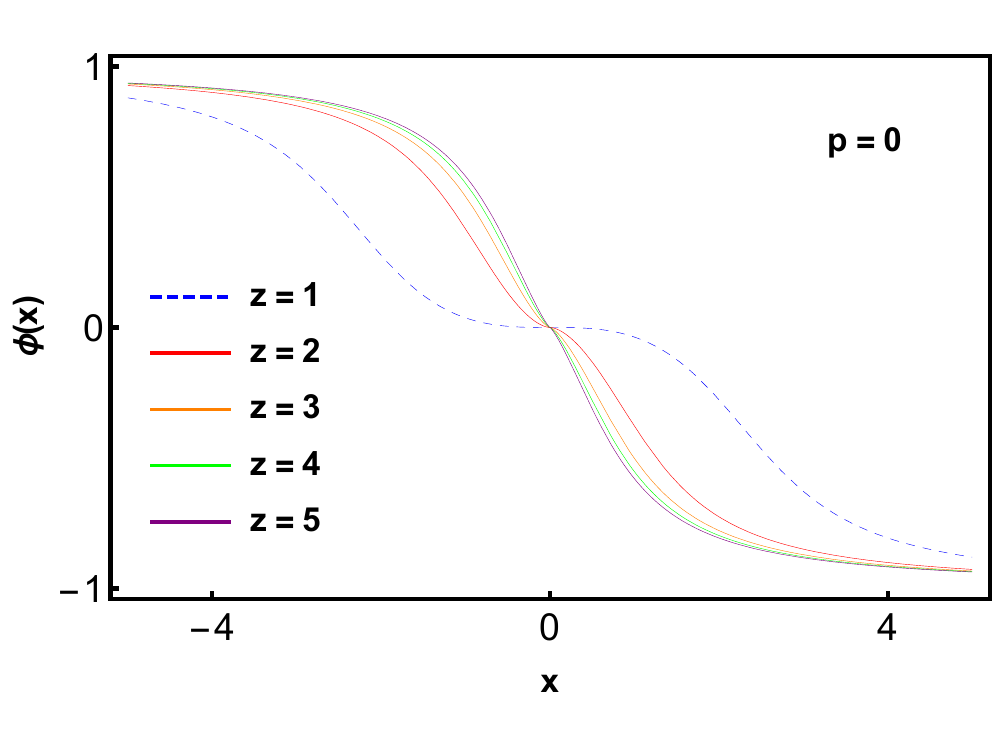}}
  \caption{Field solution $\phi(x)$ vs. $x$ without asymmetry (i.e., $p=0$) and varying $z$. For all cases, we adopt $\lambda=\nu=q=1$.}
  \label{fig3}
\end{figure}

\begin{figure}[!ht]
  \centering
  \subfigure[Asymmetric double-kink-like solutions]{\includegraphics[height=5.5cm,width=5.4cm]{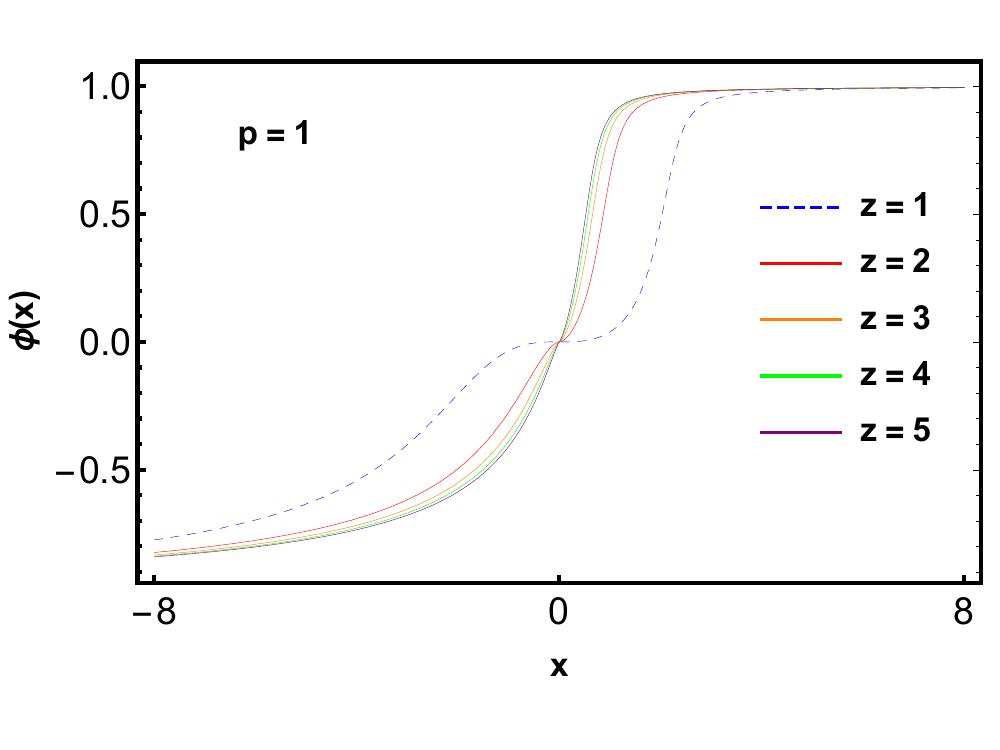}}\hspace{1cm}
   \subfigure[Asymmetric double-antikink-like solutions.]{\includegraphics[height=5.5cm,width=5.4cm]{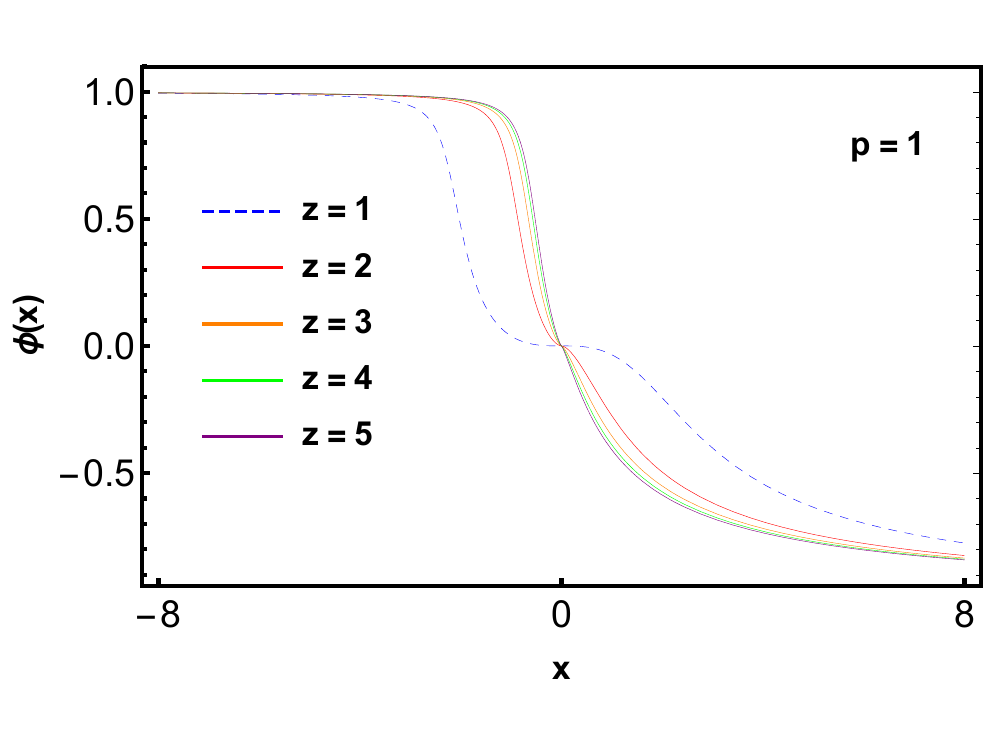}}
  \caption{Field solution $\phi(x)$ vs. $x$ with left-handed asymmetry (i.e., $p=1$) and varying $z$. For all cases, we use $\lambda=\nu=q=1$.}
  \label{fig4}
\end{figure}

\begin{figure}[!ht]
  \centering
  \subfigure[Asymmetric double-kink-like solutions.]{\includegraphics[height=5.5cm,width=5.4cm]{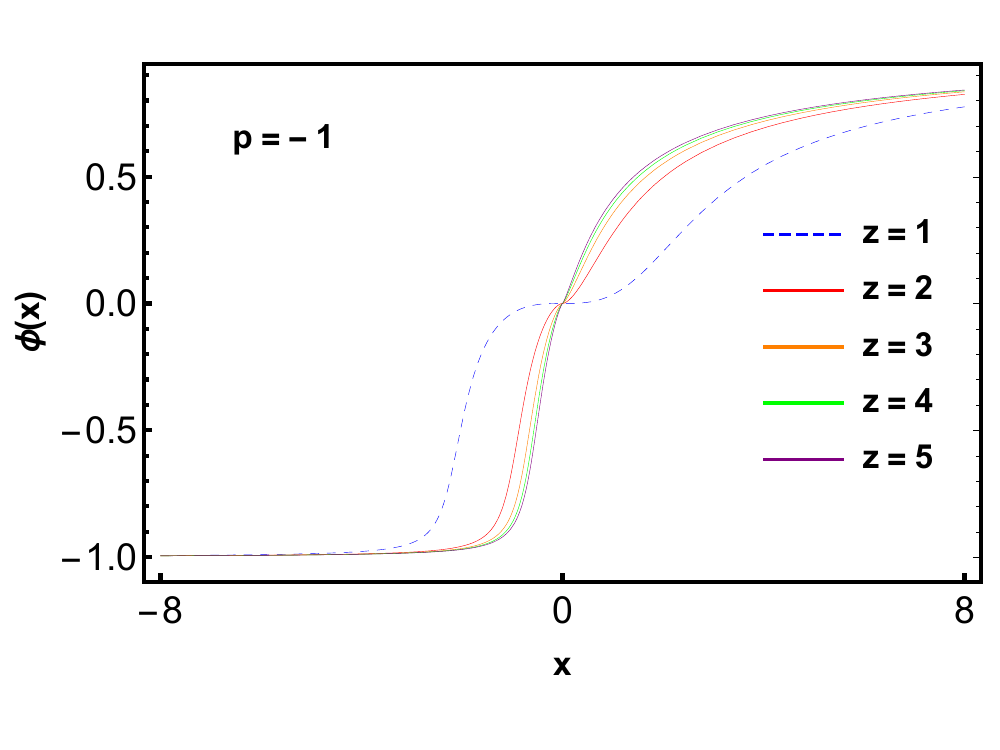}}\hspace{1cm}
   \subfigure[Asymmetric double-antikink-like solutions.]{\includegraphics[height=5.5cm,width=5.4cm]{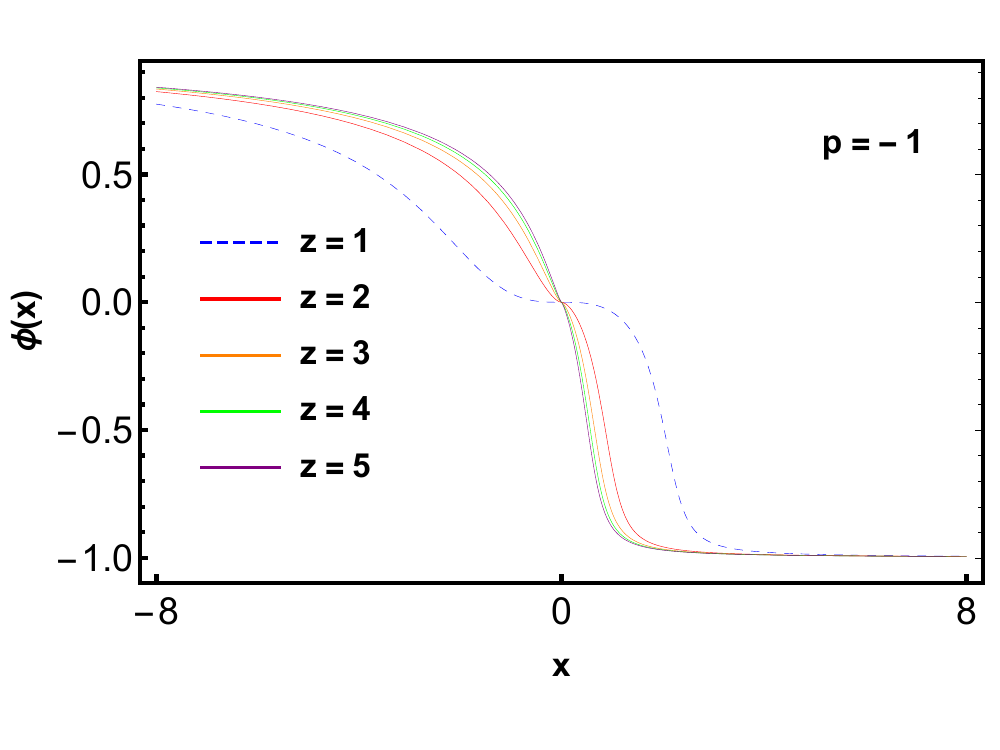}}
  \caption{Field solution $\phi(x)$ vs. $x$ with right-handed asymmetry (i.e., $p=1$) and varying $z$. For all cases, we use $\lambda=\nu=q=1$.}
  \label{fig5}
\end{figure}

\begin{figure}[!ht]
  \centering
  \subfigure[Asymmetric double-kink-like solutions.]{\includegraphics[height=5.5cm,width=5.4cm]{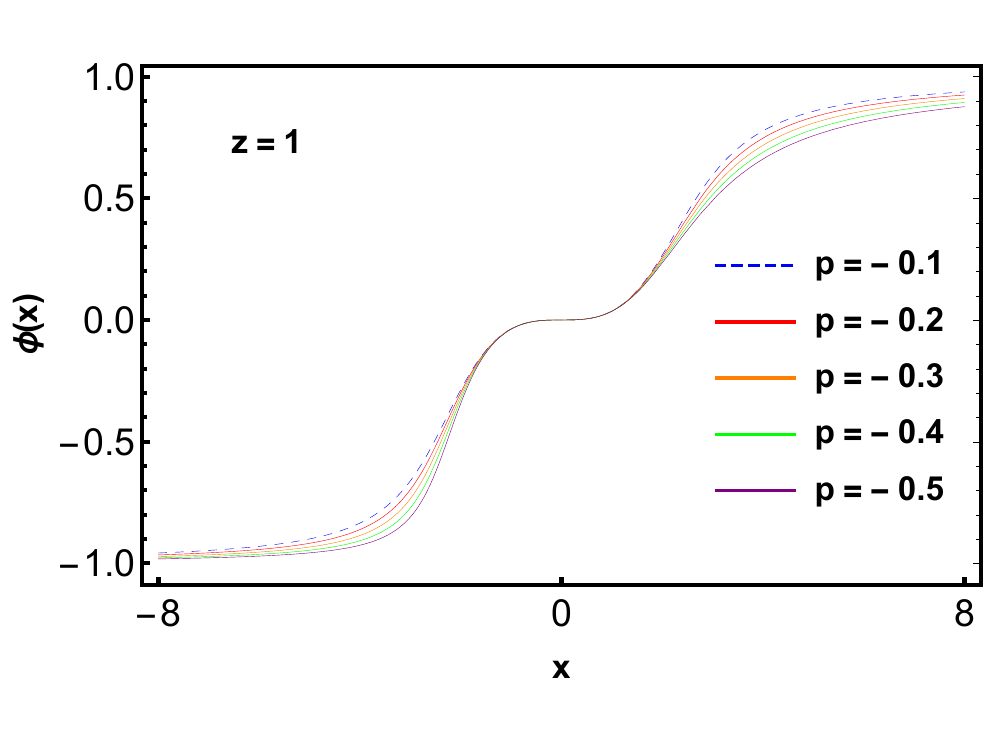}}\hspace{1cm}
   \subfigure[Asymmetric double-antikink-like solutions.]{\includegraphics[height=5.5cm,width=5.4cm]{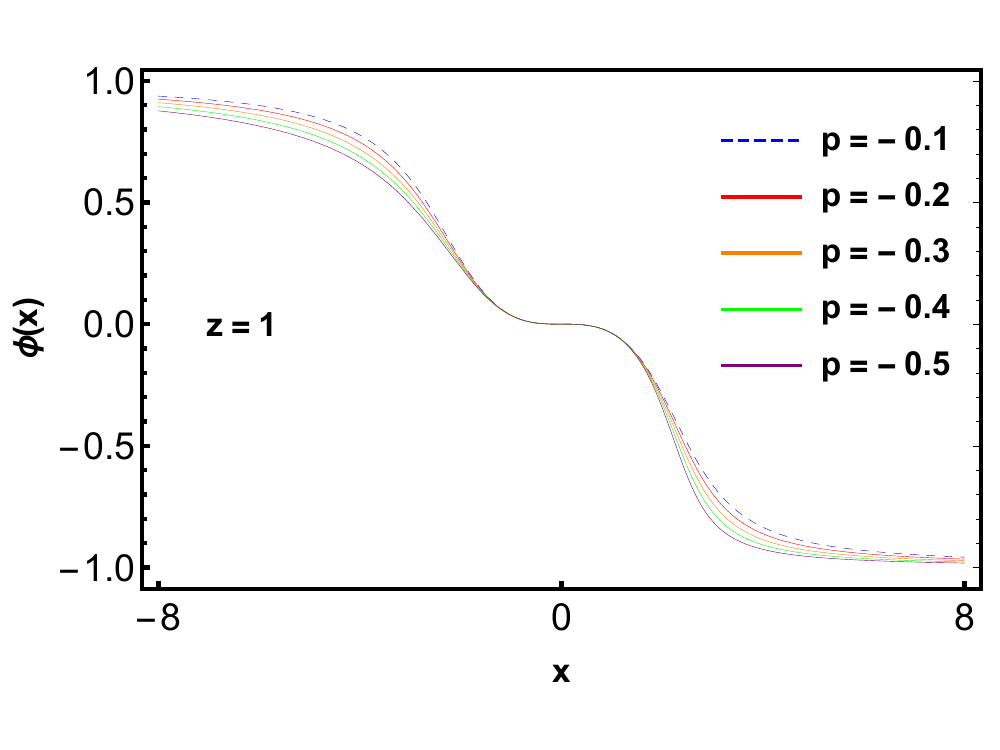}}
  \caption{Field solution $\phi(x)$ vs. $x$ with right-handed asymmetry keeping $z=1$ and varying $p<1$. For all cases, we assume $\lambda=\nu=q=1$.}
  \label{fig6}
\end{figure}

\begin{figure}[!ht]
  \centering
  \subfigure[Asymmetric double-kink-like solutions.]{\includegraphics[height=5.5cm,width=5.4cm]{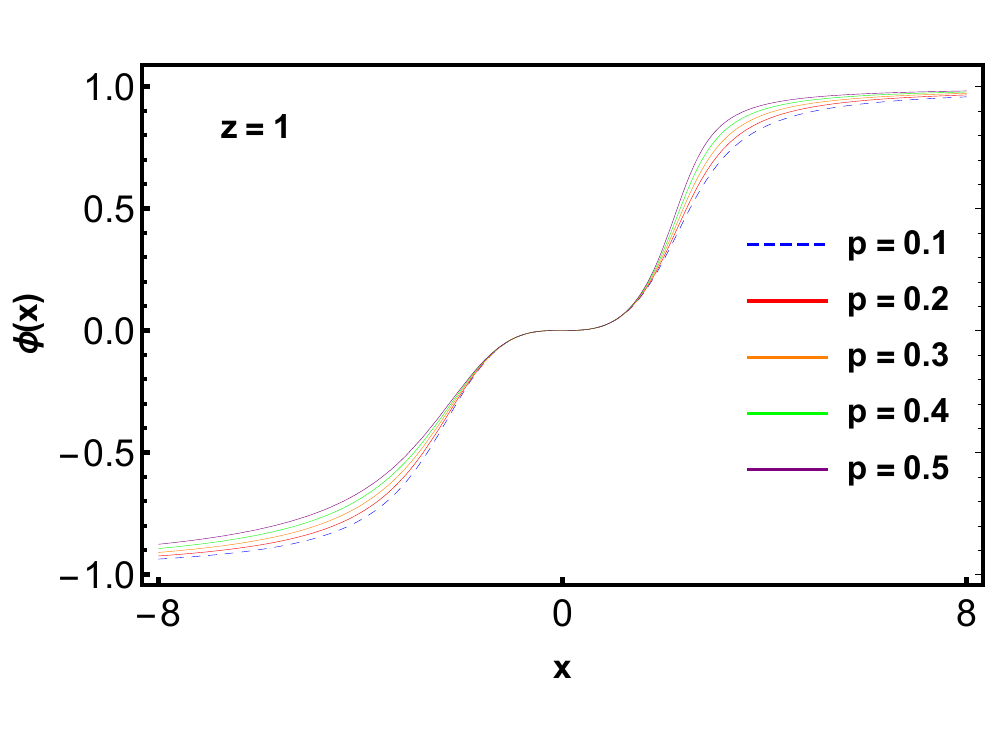}}\hspace{1cm}
   \subfigure[Asymmetric double-antikink-like solutions.]{\includegraphics[height=5.5cm,width=5.4cm]{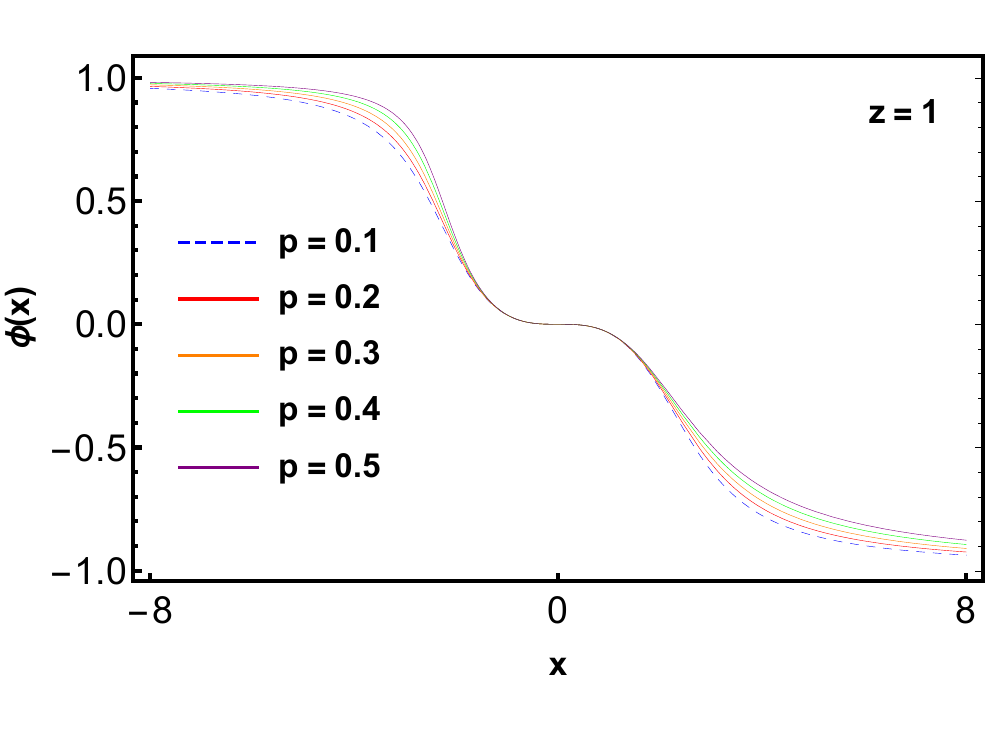}}
  \caption{Field solution $\phi(x)$ vs. $x$ with left-handed asymmetry keeping $z=1$ and varying $p>0$. For all cases, we adopt $\lambda=\nu=q=1$.}
  \label{fig7}
\end{figure}

Once the system setup is entirely defined, i.e., the generalizing function \eqref{Eq38} and the potential \eqref{Eq39} are established. We now numerically analyze the profiles of the hyperbolically modified $\phi^4$ theory. In pursuit of this purpose, it is adopted the numerical interpolation approach exposed in section \ref{sec2a} to solve Eq. \eqref{Eq40} numerically\footnote{Furthermore, the topological boundary condition $\phi_{\pm\infty}\to \pm \nu$ describes the kink-like solutions, and $\phi_{\mp\infty}\to \pm \nu$ the antikink-like configurations.}. Naturally, this results in the numerical solutions displayed in Figs. \ref{fig3} -- \ref{fig7}.

By examining the numerical solutions displayed in Figs. \ref{fig3} -- \ref{fig7}, one notes the appearance of deformations in the field configurations. Briefly, this deformation describes a continuous transformation from kink-like configurations into double-kink-like profiles as the change parameter $z$, and without asymmetry ($p = 0$). This result is consistent with those previously reported in the context of the $\phi^4$ theory \cite{Lima2,Lima3}\footnote{Note that this consistency stems from the fact that, $p=0$ and $z\to\infty$, the potential boils down to $V(\phi)=\frac{\lambda}{2}(\nu^2-\phi^2)^2$, which yields the solutions $\phi(x)=\pm \nu \tanh(\eta x)$ with $\eta=\nu\sqrt{\lambda}$.}. Such behavior can be confirmed by analyzing Figs. \ref{fig3}(a) and \ref{fig3}(b).

By introducing asymmetry into the theory, i.e., by assuming $p \neq 0$, we noted the appearance of asymmetric\footnote{The symmetric field configuration is a solution of the equation of motion that satisfies specific properties, such as spatial reflection ($\phi(x) = \phi(-x)$, parity); translational invariance ($\phi(x + a) = \phi(x)$), and internal symmetries ($\phi = -\phi$). Therefore, if the violation of the internal and translational symmetries, asymmetric field configurations appear, altering some physical properties, e.g., the energy density.} double-kink profiles. For instance, one can identify the appearance of a deformation exhibiting these characteristics in Figs. \ref{fig4}(a) and \ref{fig4}(b). In the specific case of $p = 1$, a left-handed asymmetry arises [see Figs. \ref{fig1}(a) and \ref{fig2}(b)], resulting from the lower localization of the vacuum at $\phi = -\nu$. That leads to kink/antikink solutions, where the kink profile interpolates between the topological sectors $[-\nu, \nu]$.

Within the range $[0, \nu]$, the configuration resembles a standard kink profile, whereas in the limit $\phi \to \nu$, a compacted behavior is noted at the top of the structure [see Fig. \ref{fig4}(a)]. The antikink solution [Fig. \ref{fig4}(b)] exhibits an analogous behavior, with a kink-like profile in $[0, \nu]$ and compactification in the sector $[-\nu, 0]$.

For $p = -1$, a similar behavior becomes notorious. However, in this case,  the behavior reflects the results $p = -1$, but with compactification at the bottom of the structure [see Figs. \ref{fig4}(a) and \ref{fig4}(b)]. That is because, for $p < 0$, the asymmetry is shifted around the vacuum at $\phi = \nu$, making it more asymmetric compared to the vacuum at $\phi = -\nu$.

Finally, we analyze the field profiles by keeping the deformation parameter ($z = 1$) and varying the asymmetry parameter ($p$) negatively [Figs. \ref{fig6}(a) and \ref{fig6}(b)] and positively [Figs. \ref{fig7}(a) and \ref{fig7}(b)]. In these cases, only double-kink-like profiles emerge. For $p < 0$, the field is mitigated in the limit $\phi \to \nu$, while for $p > 0$, this mitigation becomes pervasive as $\phi \to -\nu$.

\begin{figure}[!ht]
  \centering
  \subfigure[The case without asymmetry and varying $z$.]{\includegraphics[height=4.2cm,width=5.2cm]{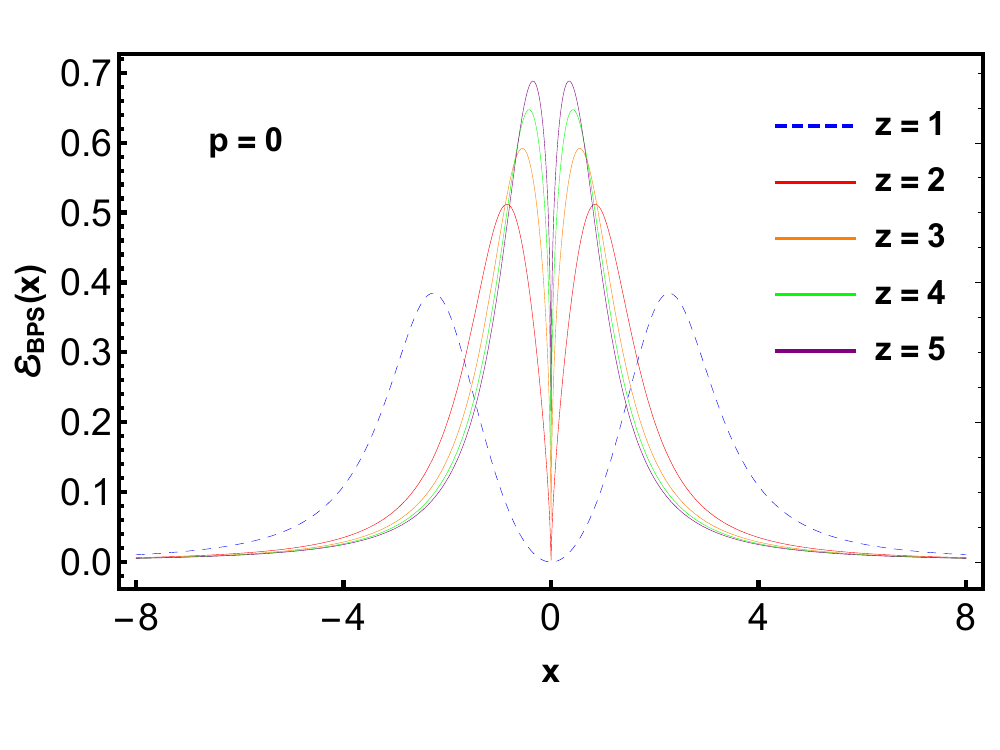}}\hfill
  \subfigure[The case with left-handed asymmetry and varying $z$.]{\includegraphics[height=4.2cm,width=5.2cm]{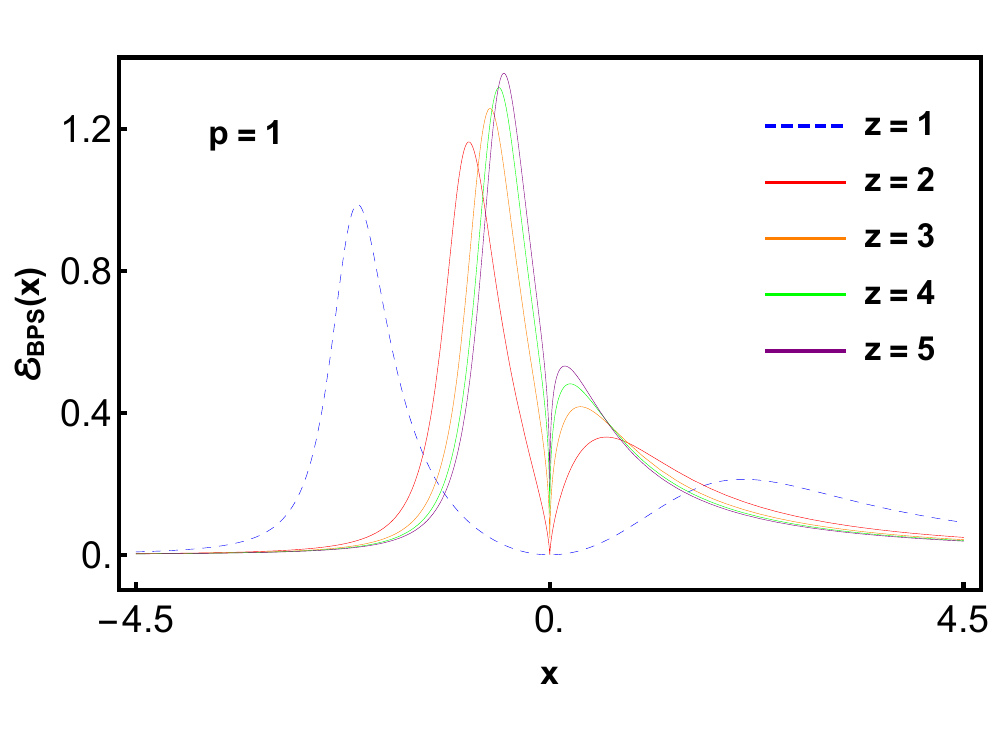}}\hfill
   \subfigure[The case with right-handed asymmetry and varying $z$.]{\includegraphics[height=4.2cm,width=5.2cm]{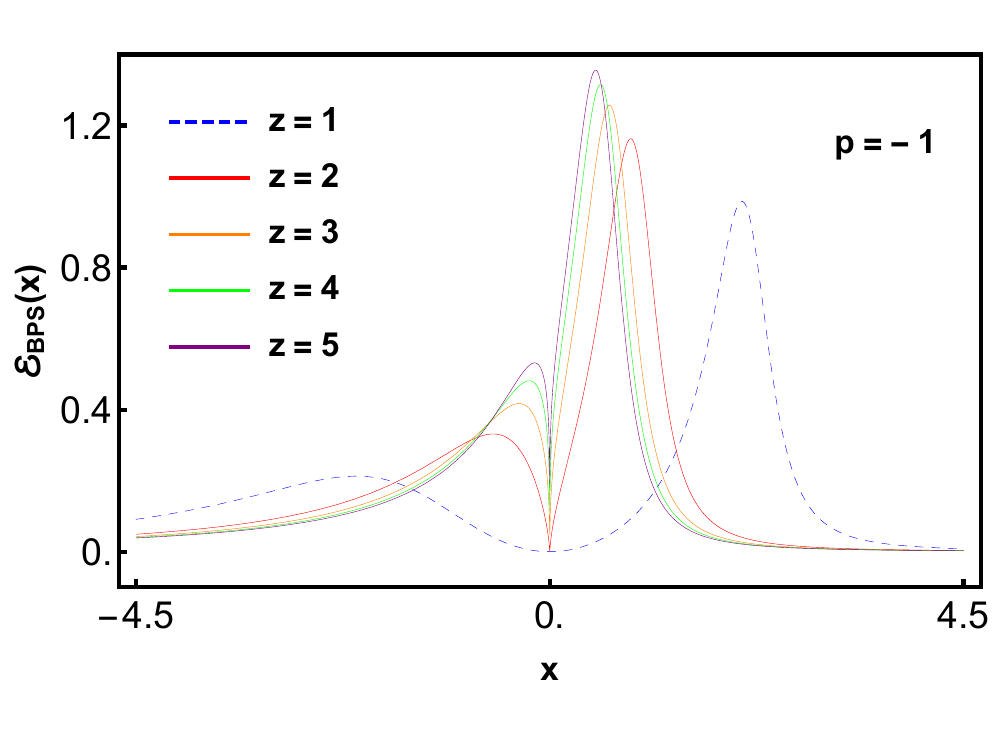}}\hfill
  \subfigure[The case with right-handed asymmetry and varying $p$.]{\includegraphics[height=4.2cm,width=5.2cm]{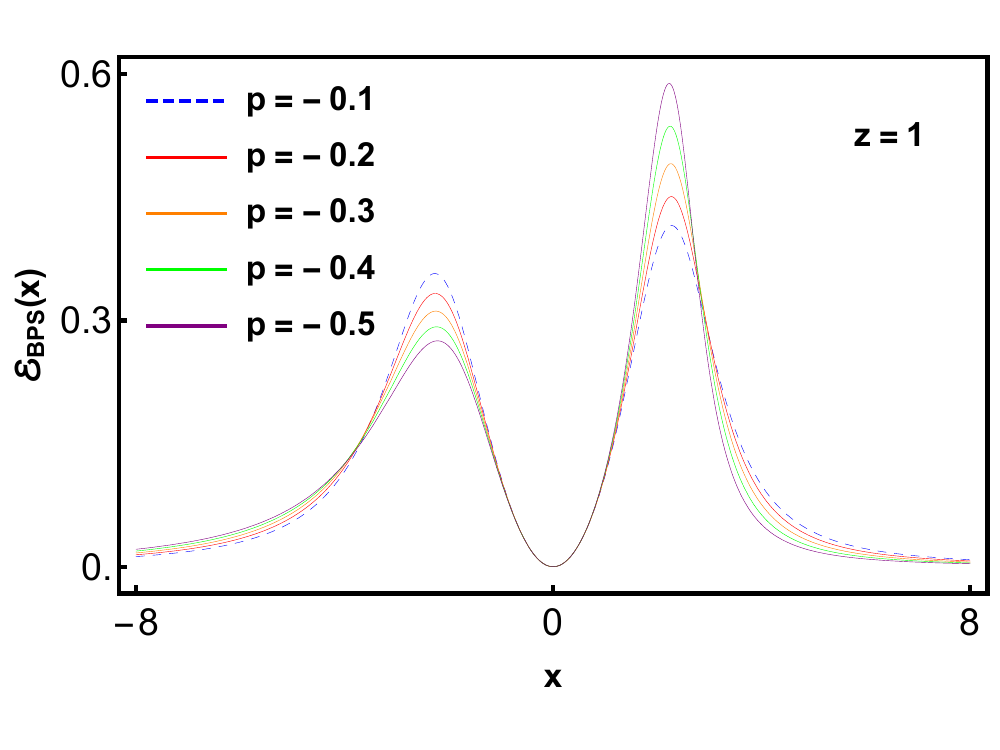}}
  \subfigure[The case with left-handed asymmetry and varying $p$.]{\includegraphics[height=4.2cm,width=5.2cm]{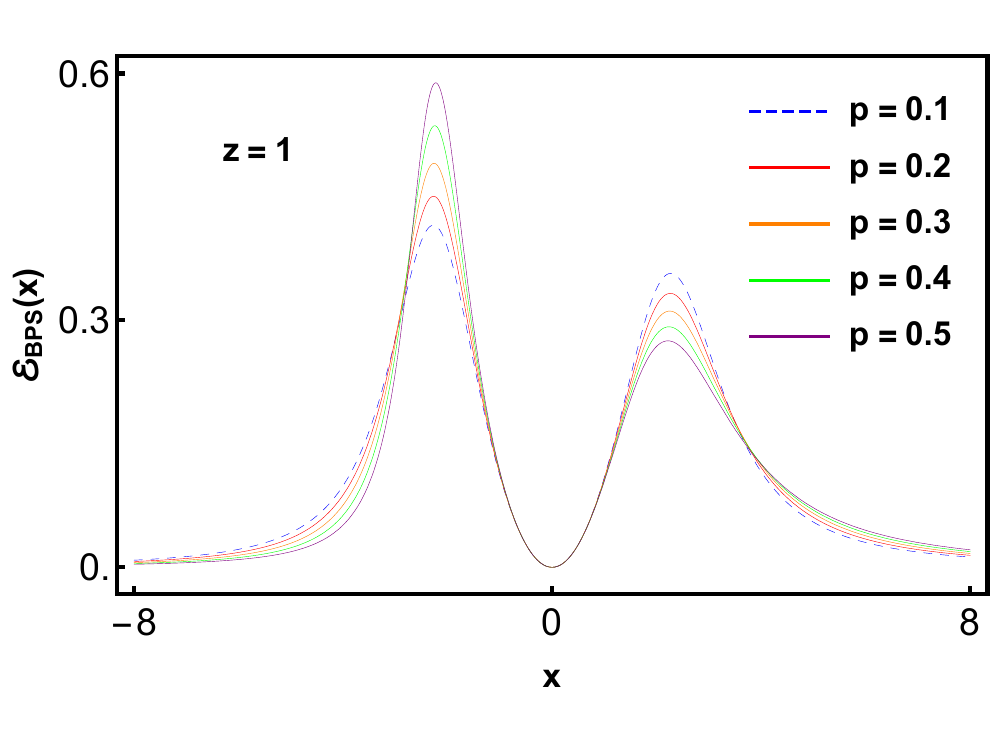}}
  \caption{BPS energy density $\mathcal{E}_{\text{BPS}}(x)$ vs. position $x$ for the kink/antikink solutions exposed in Figs. \ref{fig3}--\ref{fig7}. In all cases, we adopt $\lambda=\nu=q=1$.}  \label{fig8}
\end{figure}

Considering the numerical field solutions announced in Figs. \ref{fig3}(a)–\ref{fig7}(b), we can use Eq. \eqref{Eq12}, together with the approach presented in section \ref{sec2a}, to numerically examine the BPS energy density from the scalar field solutions\footnote{We highlight that double-kink-like and double-antikink-like configurations exhibit the same BPS energy density.}. Thus, we show the BPS energy densities in Figs. \ref{fig8}(a)–\ref{fig8}(e). These configurations confirm that the field configurations are a class of symmetric double-kink/antikink-like solutions for $p = 0$. Naturally, these fields modify when $p$ increases (or decreases), i.e., this change alters the amplitude and spatial localization. One highlights that the asymmetry becomes persistent when $p \neq 0$, while the deformation appears when $z \neq 0$.

Taking into account the Eq. \eqref{Eq18} and the topological conditions $\phi_{\pm\infty} \to \pm \nu$ for kink-like solutions and $\phi_{\mp\infty} \to \pm \nu$ for antikink-like solutions, it follows that the BPS energy of the deformed field configurations \cite{Lima1,Lima2,Lima3} is 
\begin{align}
\label{Eq41}
    \text{E}_{\text{BPS}}=\pm\frac{4}{3}\sqrt{\lambda}\nu^3.
\end{align}
Thus, we observed that for a constant value of the parameter $\lambda$, there exists a family of solutions corresponding to different values of $\nu$, implying the existence of infinitely many degenerate solutions within this theory.

\subsubsection{Linear stability}\label{sec2b3}

\begin{figure}[!ht]
  \centering
  \includegraphics[height=6cm,width=5.9cm]{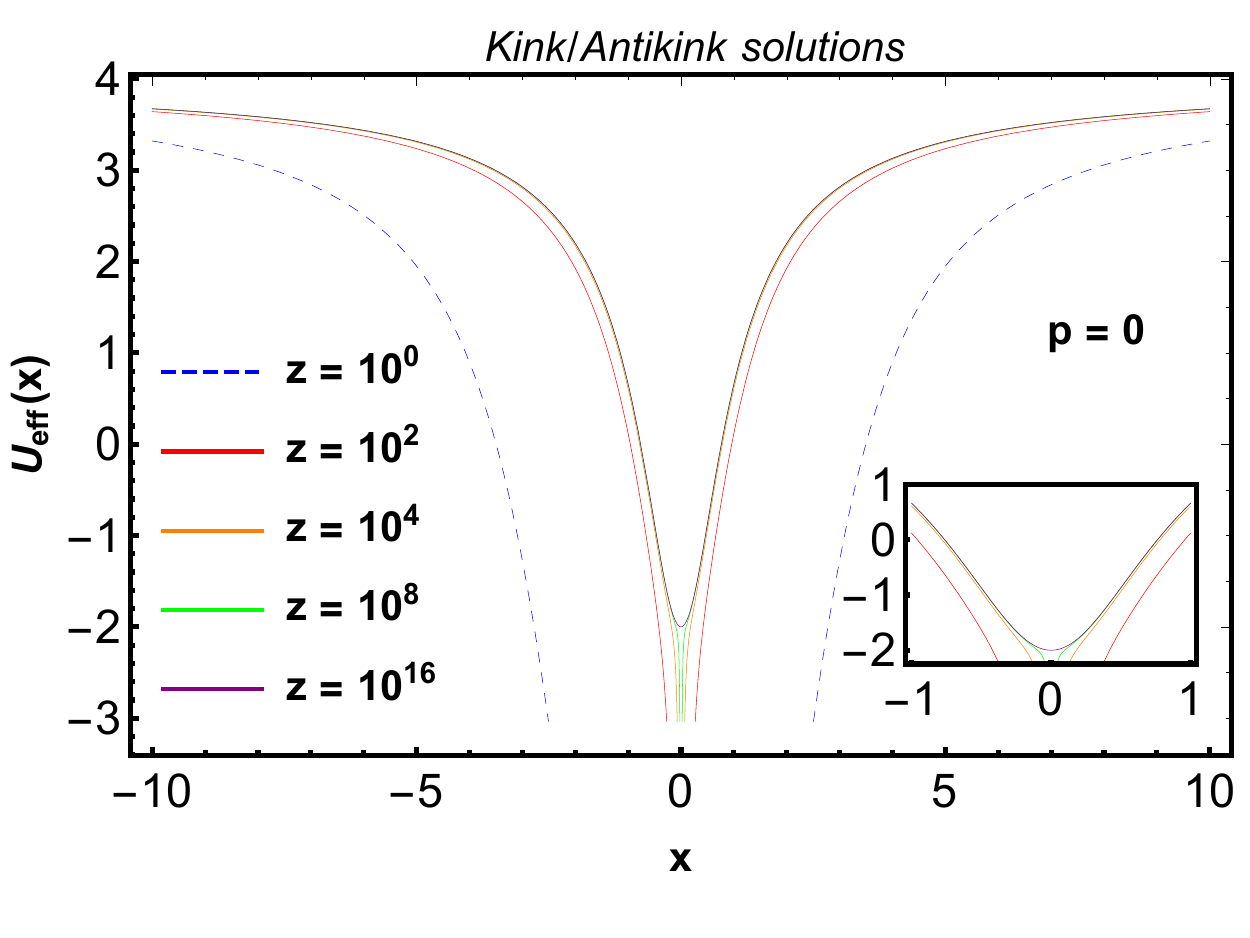}
  \caption{The effective stability potential $U_{\text{eff}}(x)$ vs. $x$ for the symmetric case (i.e., $p=0$) and varying $z$. In all cases, we assume $\lambda=\nu=q=1$.}  \label{fig9}
\end{figure}

\begin{figure}[!ht]
  \centering
  \subfigure[Effective potential $U_{\text{eff}}$ corresponding to asymmetric double-kink solutions.]{\includegraphics[height=5.5cm,width=5.4cm]{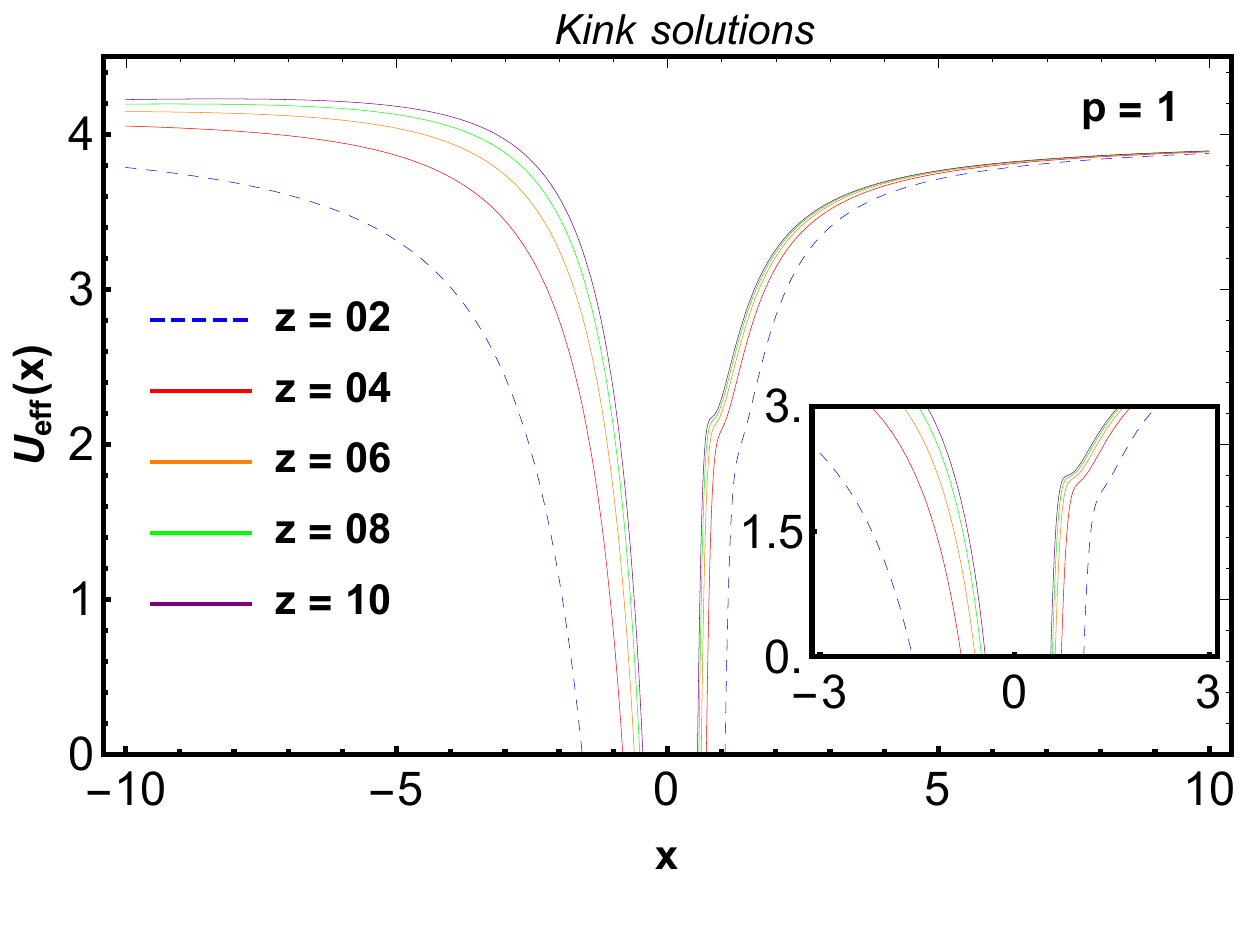}}\hspace{1cm}
   \subfigure[Effective potential $U_{\text{eff}}$ corresponding to asymmetric double-antikink solutions.]{\includegraphics[height=5.5cm,width=5.4cm]{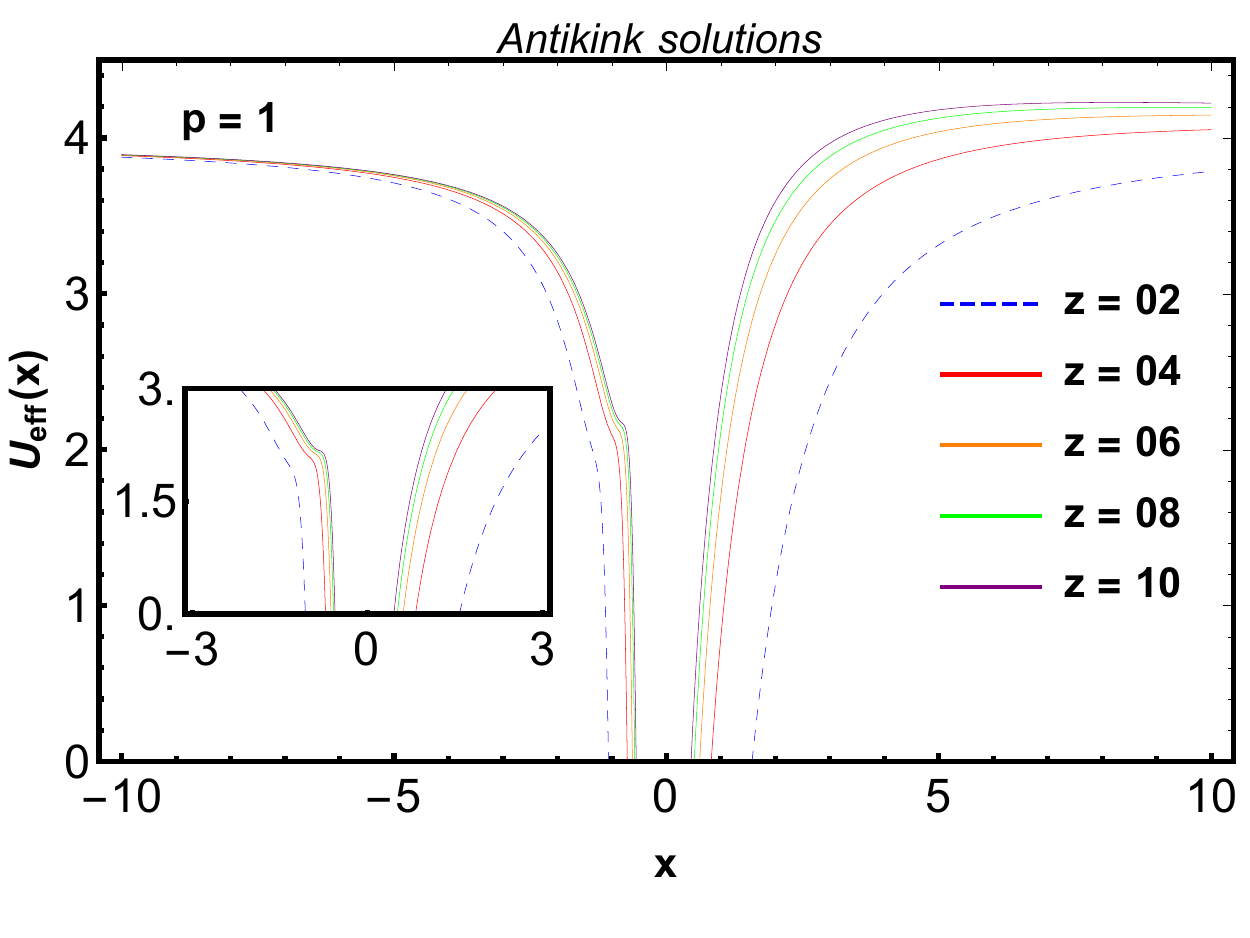}}\hfill
  \caption{The effective stability potential $U_{\text{eff}}(x)$ vs. $x$ with left-handed asymmetry (i.e., $p=1$) and varying $z$. In all cases, we assume $\lambda=\nu=q=1$.}  \label{fig10}
\end{figure}

\begin{figure}[!ht]
  \centering
  \subfigure[Effective potential $U_{\text{eff}}$ corresponding to asymmetric double-kink solutions.]{\includegraphics[height=5.5cm,width=5.4cm]{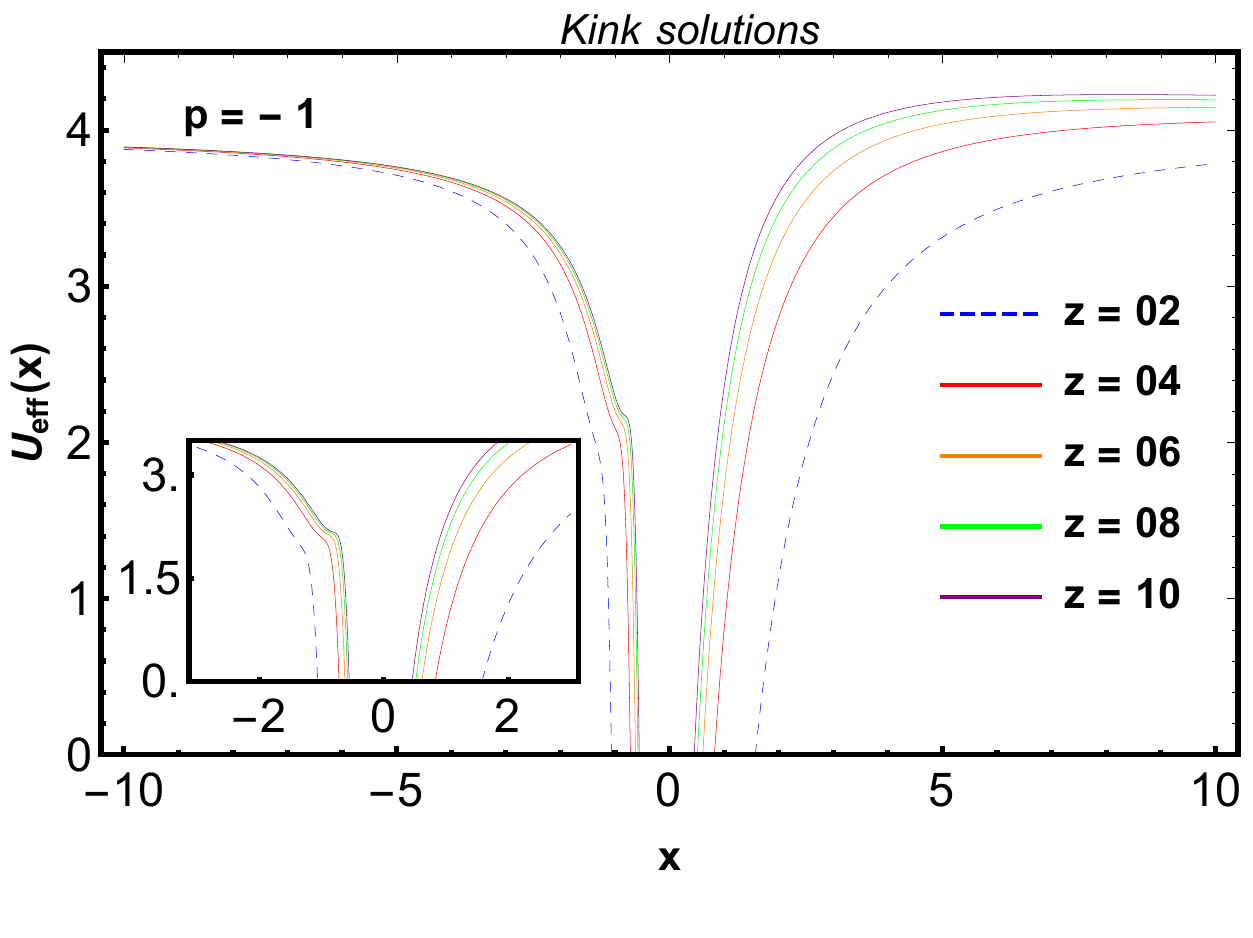}}\hspace{1cm}
   \subfigure[Effective potential $U_{\text{eff}}$ corresponding to asymmetric double-antikink solutions.]{\includegraphics[height=5.5cm,width=5.4cm]{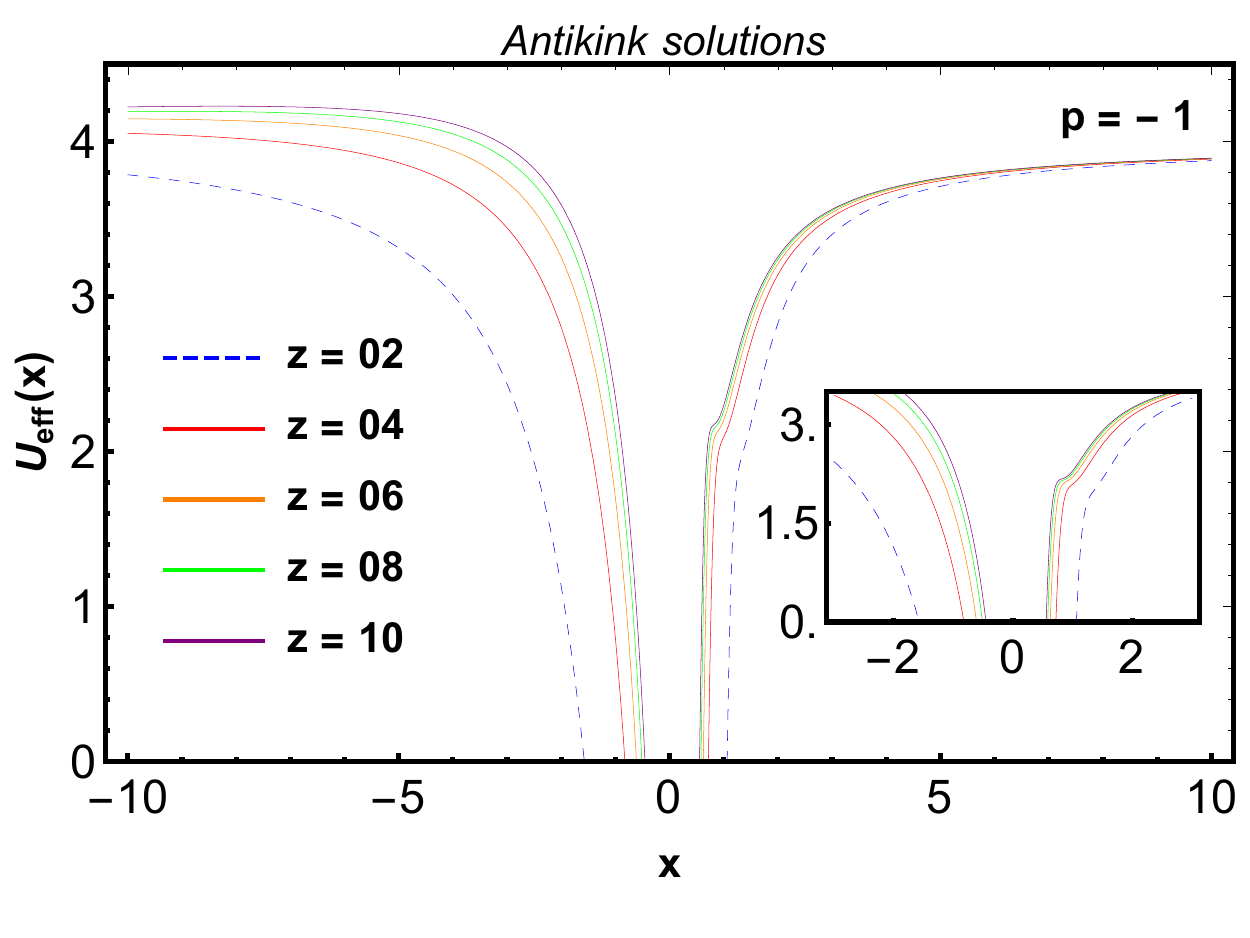}}\hfill
  \caption{The effective stability potential $U_{\text{eff}}(x)$ vs. $x$ with right-handed asymmetry (i.e., $p=1$) and varying $z$. In all cases, we assume $\lambda=\nu=q=1$.}  \label{fig11}
\end{figure}

\begin{figure}[!ht]
  \centering
  \subfigure[Effective potential $U_{\text{eff}}$ corresponding to asymmetric double-kink solutions.]{\includegraphics[height=5.5cm,width=5.4cm]{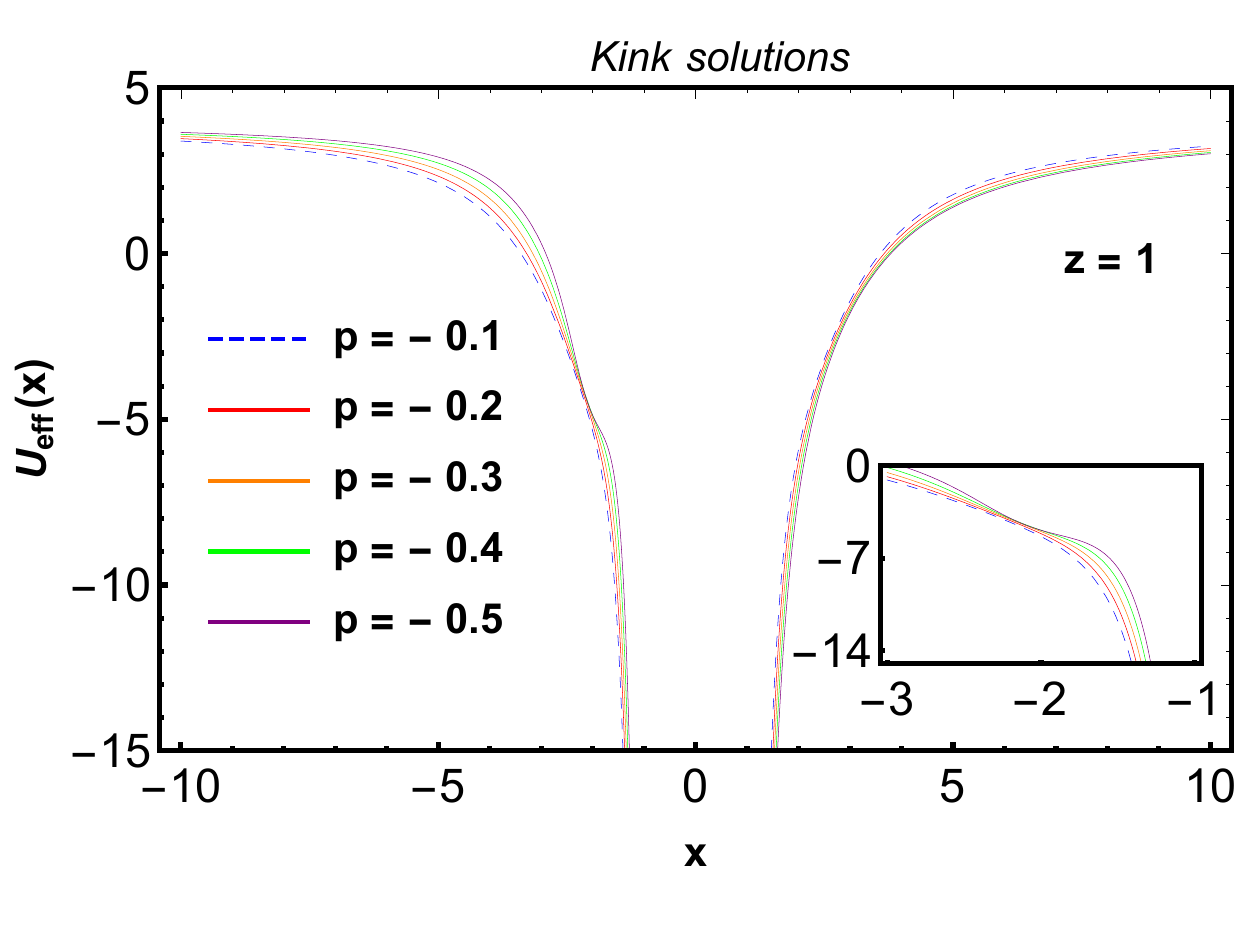}}\hspace{1cm}
   \subfigure[Effective potential $U_{\text{eff}}$ corresponding to asymmetric double-antikink solutions.]{\includegraphics[height=5.5cm,width=5.4cm]{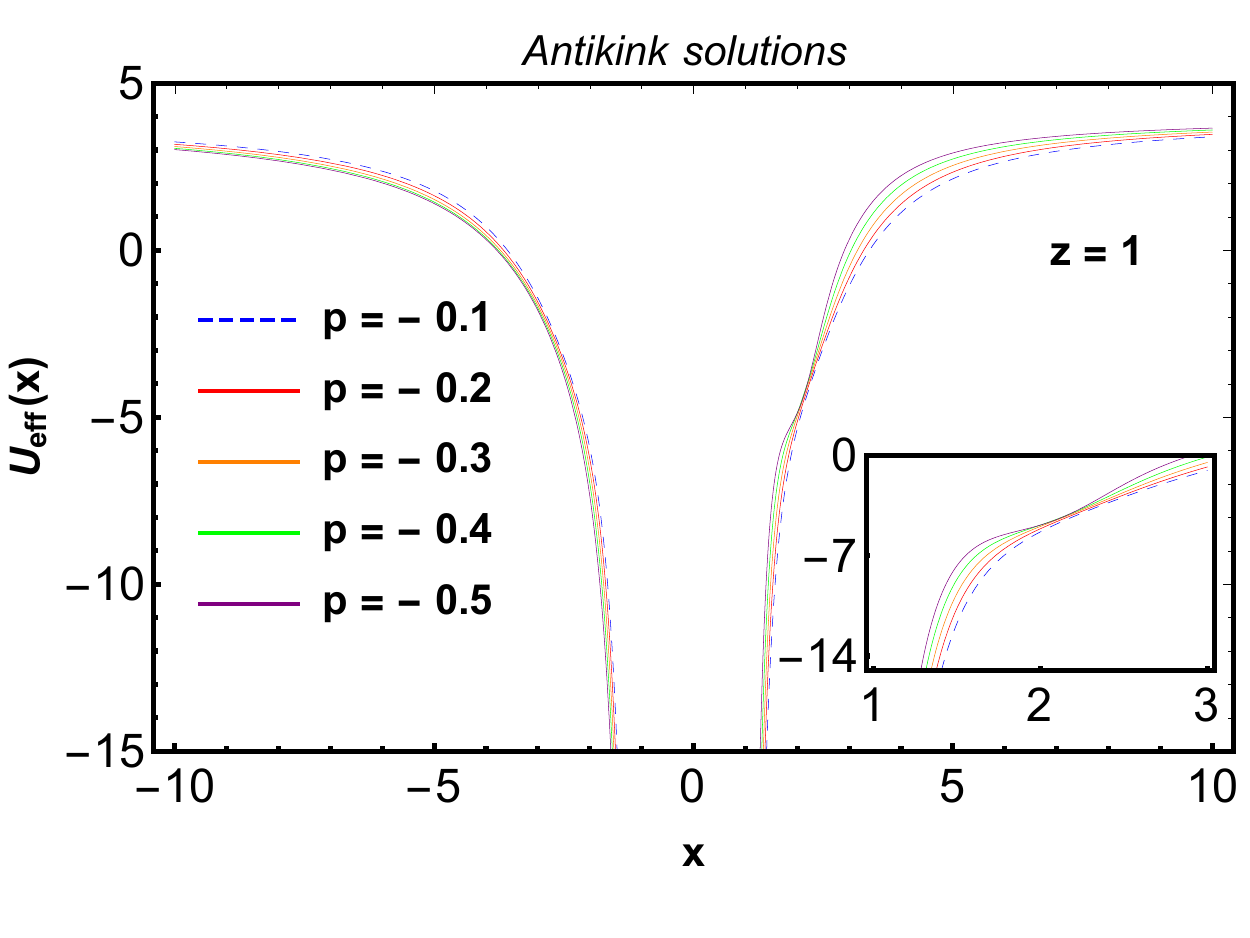}}\hfill
  \caption{The effective stability potential $U_{\text{eff}}(x)$ vs. $x$ with right-handed asymmetry (i.e., $p<0$) and keeping $z=1$. In all cases, we adopt $\lambda=\nu=q=1$.}  \label{fig12}
\end{figure}

\begin{figure}[!ht]
  \centering
  \subfigure[Effective potential $U_{\text{eff}}$ corresponding to asymmetric double-kink solutions.]{\includegraphics[height=5.5cm,width=5.4cm]{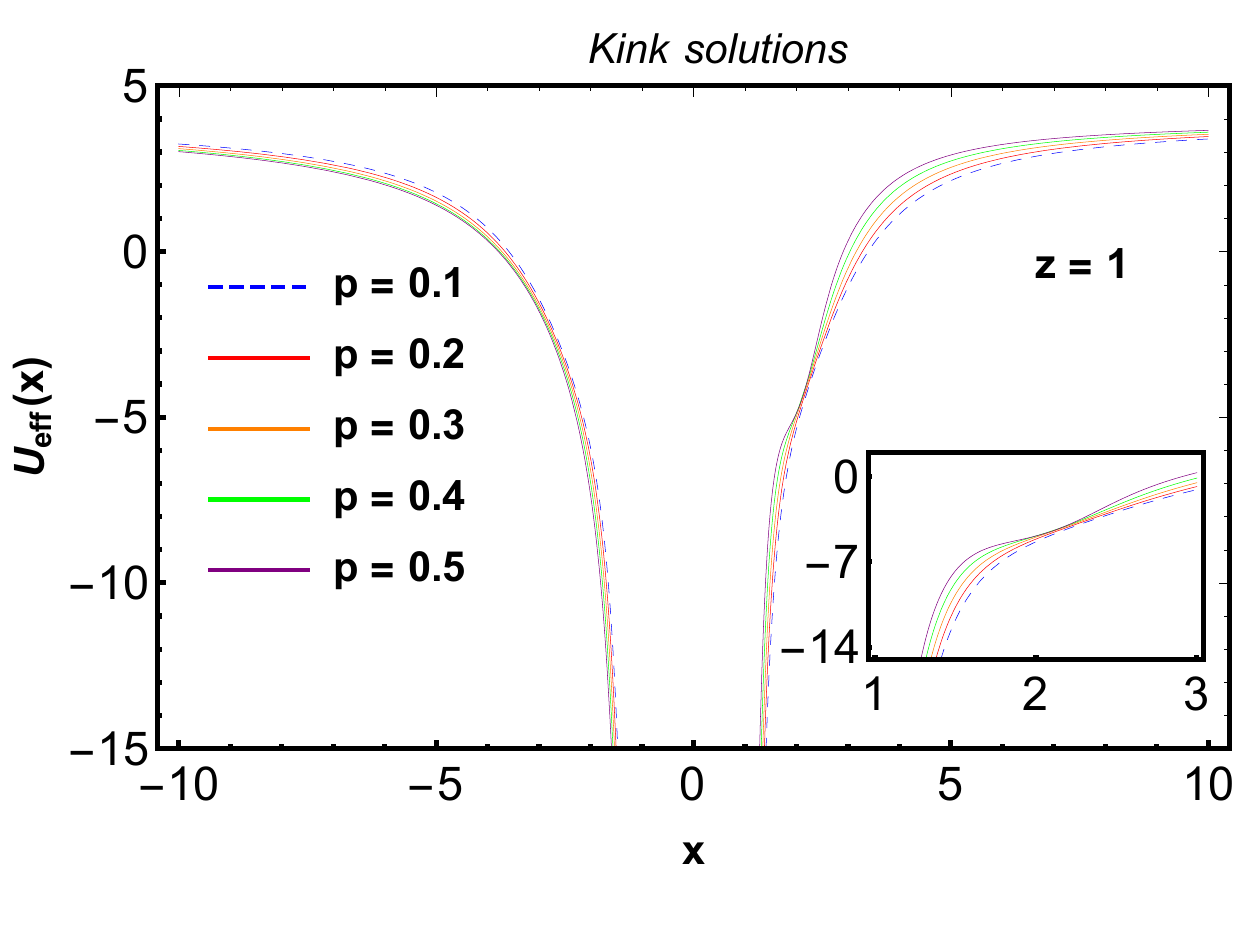}}\hspace{1cm}
   \subfigure[Effective potential $U_{\text{eff}}$ corresponding to asymmetric double-antikink solutions.]{\includegraphics[height=5.5cm,width=5.4cm]{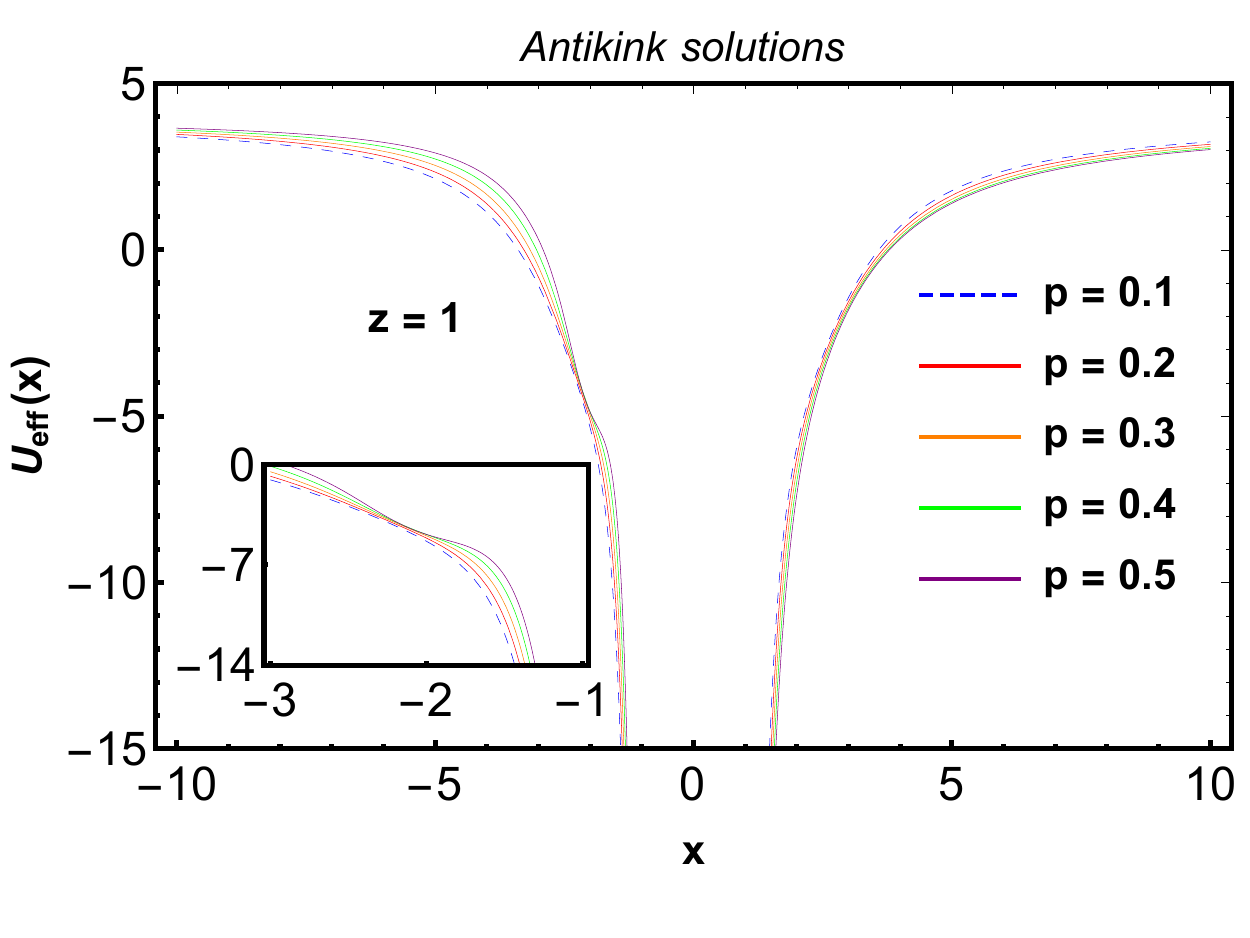}}\hfill
  \caption{The effective stability potential $U_{\text{eff}}(x)$ vs. $x$ with left-handed asymmetry (i.e., $p>0$) and keeping $z=1$. In all cases, we adopt $\lambda=\nu=q=1$.}  \label{fig13}
\end{figure}

Hereafter, we are prepared to analyze the linear stability of the solutions presented in section \ref{sec2b2}. For this purpose, we consider the numerical solutions exposed in Figs. \ref{fig3}(a)–\ref{fig7}(b), along with the effective potential defined in Eq. \eqref{Eq29} and the numerical method described in section \ref{sec2a}. These considerations allow us to obtain the numerical profiles of the stability potentials \eqref{Eq29}. We exposed the numerical behavior of the linear stability in Figs. \ref{fig9}–\ref{fig13}(b).

It is essential to highlight that in the limit $p\to 0$ and $z\to \infty$, one arrives at
\begin{align}
    V(\phi)\approx\frac{\lambda}{2}(\nu^2-\phi^2)^2,
\end{align}
i.e., we recovered the standard $\phi^4$ theory, which its solutions are $\phi(x)=\pm \nu\tanh(m x)$ where $m=\nu \sqrt{\lambda}$. Thus, when $p = 0$ and $z \to \infty$, the effective potential concerning the kink/antikink configurations is $U_{\text{eff}}\approx-2\lambda\nu^2[1-3\tanh^2(m x)]$, in which gives rise to the translational mode ($\omega = 0$) given by $\eta_0(x) = m\nu\, \text{sech}^2(mx)$.Therefore, in the usual $\phi^4$ theory, the translational mode resembles the Gaussian-like profile and is described by a Pöschl–Teller-like potential, whose maximum value occurs at the center of the potential well, namely, $U_{\text{eff}}^{\text{max}}(x)=4\lambda\nu^2$ for $\lambda,\,\nu\in\mathbb{R}$. Therefore, in this case, one notes that the translational symmetry is manifest, allowing a spatial displacement of the structures without any energy cost \cite{Vachaspati}.

Conversely, for more general cases involving asymmetry ($p \neq 0$) and deformation ($z > 0$), we consider the topological solutions introduced in Section \ref{sec2b2}. Accordingly, by adopting the numerical field solutions [Figs. \ref{fig3}(a) -- \ref{fig7}(b)] with Eq. \eqref{Eq29}, we compute the stability potential numerically for $p \neq 0$ and $z \neq 0$. We displayed the corresponding effective stability potential in Figs. \ref{fig9} -- \ref{fig13}(b).

By inspecting the numerical results for the stability potential [Figs. \ref{fig9} -- \ref{fig13}(b)], one notes, in the limit $p\to 0$ and $z\to \infty$, the effective stability potential assumes the standard behavior from the $\phi^4$ theory. Therefore, the stability potential converges to a P\"{o}schl–Teller-like profile when $p = 0$ and the deformation parameter increases (see Fig. \ref{fig9}). However, even in the absence of asymmetry ($p = 0$), when the deformation parameter takes small values, i.e., when double-kink-like configurations emerge, the stability potential resembles a Dirac delta-like well, i.e., $U_{\text{eff}} \propto -\delta(x)$. According to Ref. \cite{Griffiths}, at least one bound state should exist for the corresponding Schr\"{o}dinger-like equation. Naturally, in this conjecture, the presence of the translational mode ($\omega = 0$) is observed.

Moreover, when asymmetry becomes manifest ($p\neq 0$), the potential wells still diverge at the origin, which, as shown in Ref. \cite{Griffiths}, ensures the persistence of the translational mode. Nevertheless, in this context, the effective stability potential becomes asymmetric in the asymptotic regions; i.e., the value of $U_{\text{eff}}$ differs in the spacial limits. For instance, in the case of left-handed asymmetry ($p>0$), it is found that $U_{\text{eff}}(x\to +\infty)>U_{\text{eff}}(x\to -\infty)$ for kink-like (or double-kink) solutions, whereas for ($p<1$), the opposite occurs, viz., $U_{\text{eff}}(x\to +\infty)<U_{\text{eff}}(x\to -\infty)$ [see Figs. \ref{fig10}(a)–\ref{fig11}(b)]. Therefore, in the case of left-handed asymmetry, a particle moving from right to left experiences a different interaction from that of a particle moving in the opposite direction, from left to right [Figs. \ref{fig10}(a)–\ref{fig11}(b)]. Finally, it is worth noting that as the deformation parameter varies, the solutions approach the double-kink profile. This behavior provides a recurring ripple point that emerges in all curves of the effective potential, in the cases $p \neq 0$ and $z \ll \infty$ [see Figs. \ref{fig12}(a)–\ref{fig13}(b)]. This feature is located approximately within the range $\vert x\vert\in [0.9, 1.5]$.

\begin{figure}[!ht]
  \centering
  \includegraphics[height=6cm,width=5.9cm]{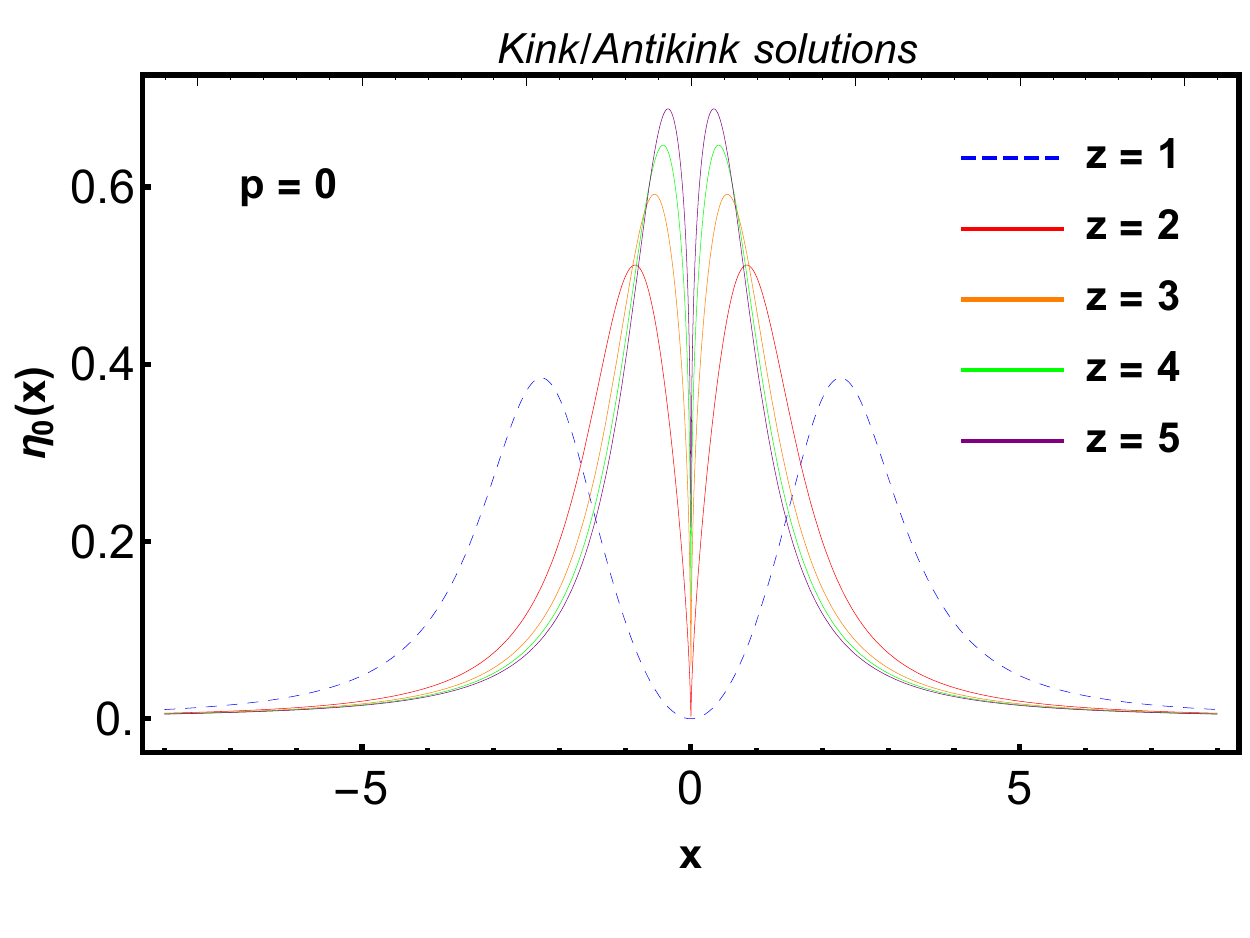}
  \caption{Translational mode $\eta_{0}(x)$ in term of $x$ for the symmetric case (i.e., $p=0$) and varying $z$. In all cases, we consider $\lambda=\nu=q=1$.}  \label{fig14}
\end{figure}

\begin{figure}[!ht]
  \centering
  \subfigure[$\eta_0(x)$ associated with asymmetric double-kink solutions.]{\includegraphics[height=5.5cm,width=5.4cm]{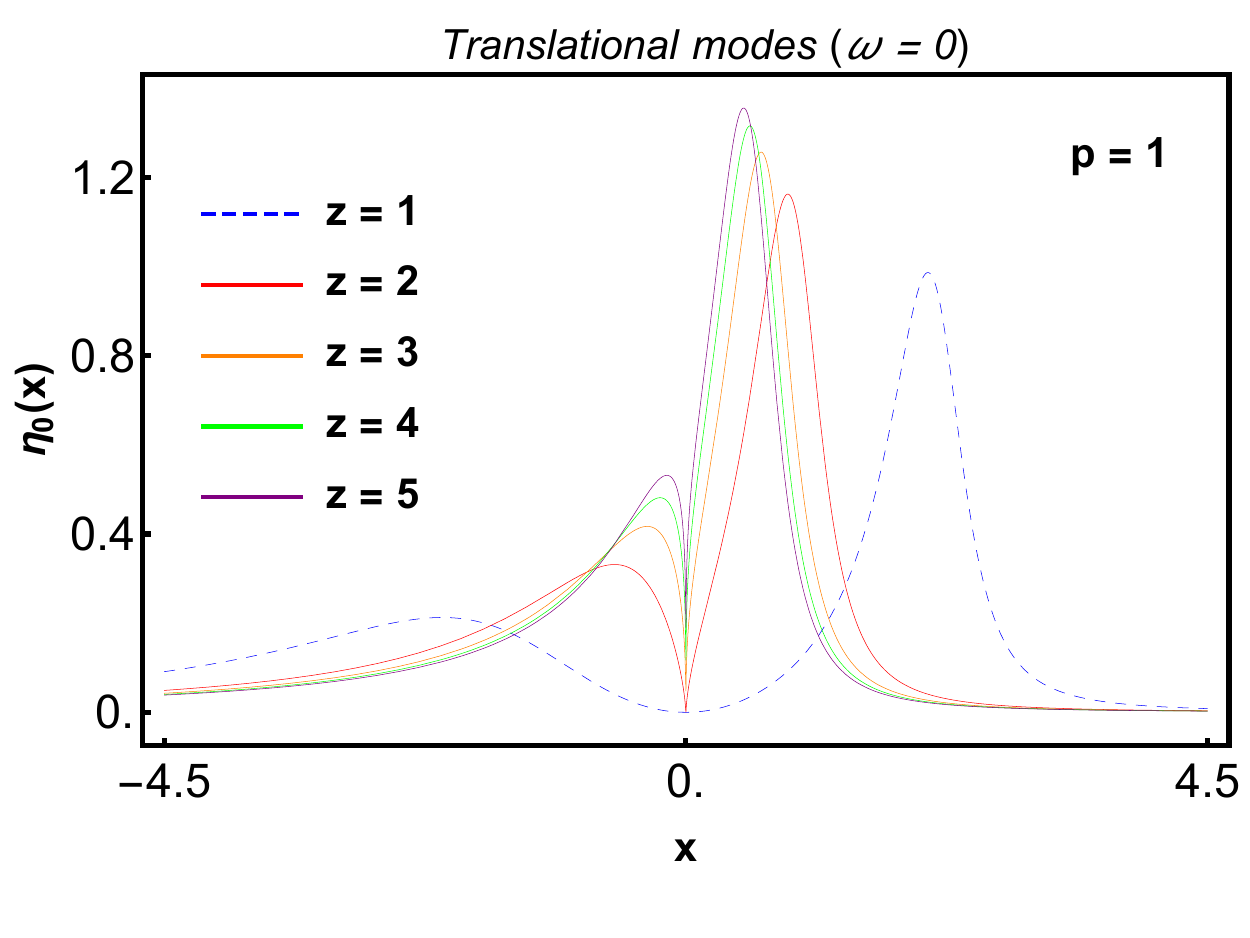}}\hspace{1cm}
   \subfigure[$\eta_0(x)$ associated with asymmetric double-antikink solutions.]{\includegraphics[height=5.5cm,width=5.4cm]{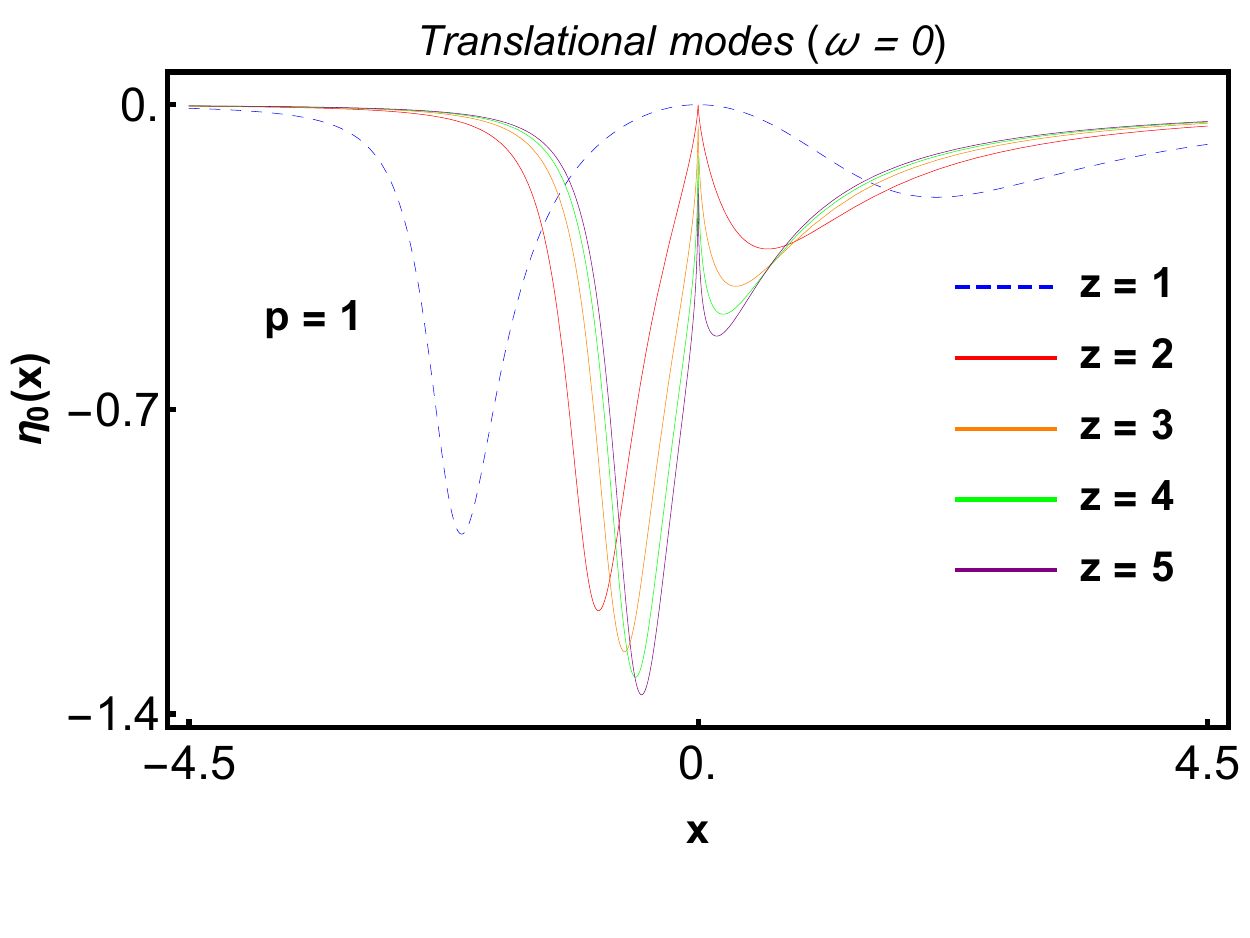}}\hfill
  \caption{Translational modes $\eta_0(x)$ vs. $x$ for left-handed asymmetry (i.e., $p=1$) and varying $z$. In all cases, we adopt $\lambda=\nu=q=1$.}  \label{fig15}
\end{figure}

\begin{figure}[!ht]
  \centering
  \subfigure[$\eta_0(x)$ associated with asymmetric double-kink solutions.]{\includegraphics[height=5.5cm,width=5.4cm]{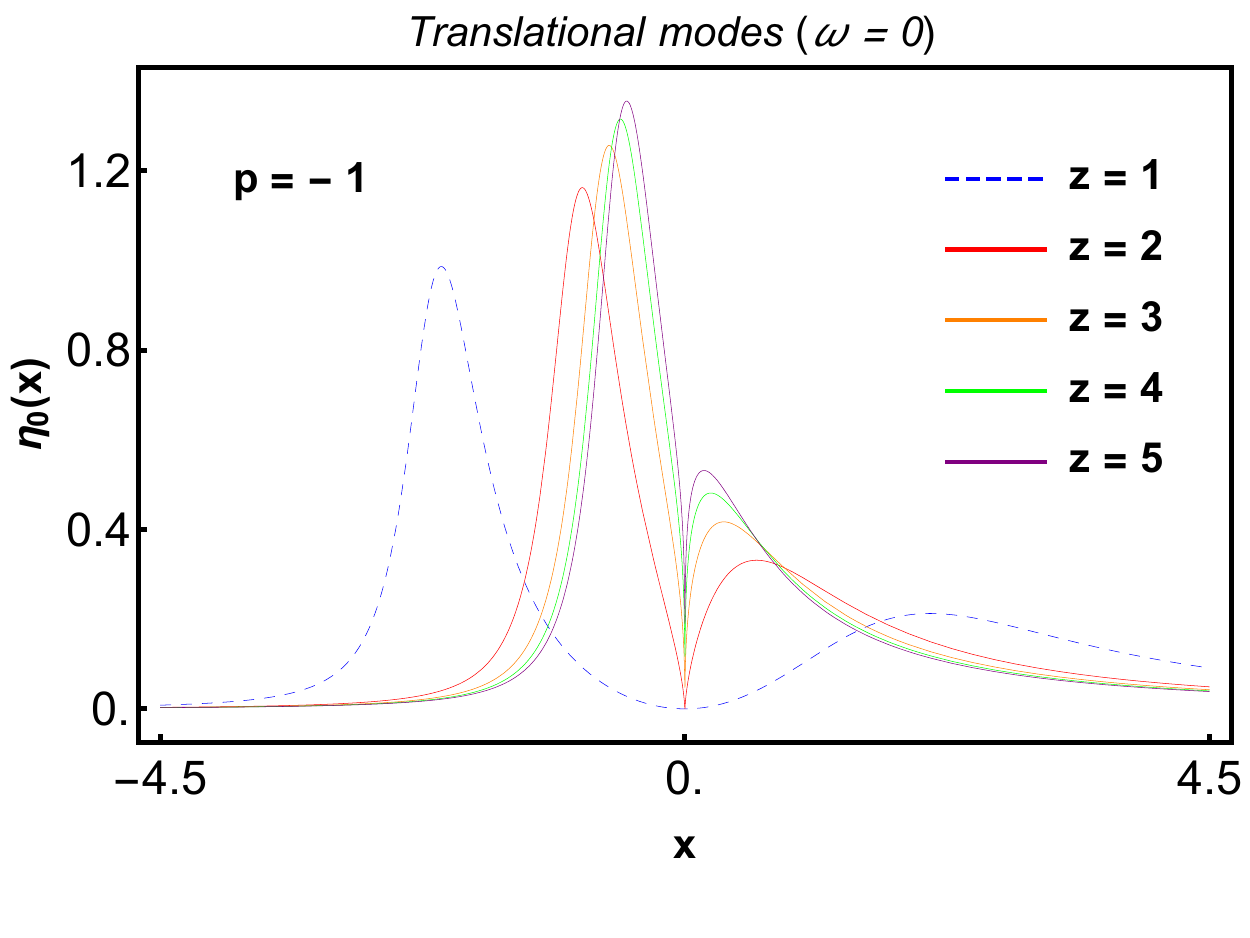}}\hspace{1cm}
   \subfigure[$\eta_0(x)$ associated with asymmetric double-antikink solutions.]{\includegraphics[height=5.5cm,width=5.4cm]{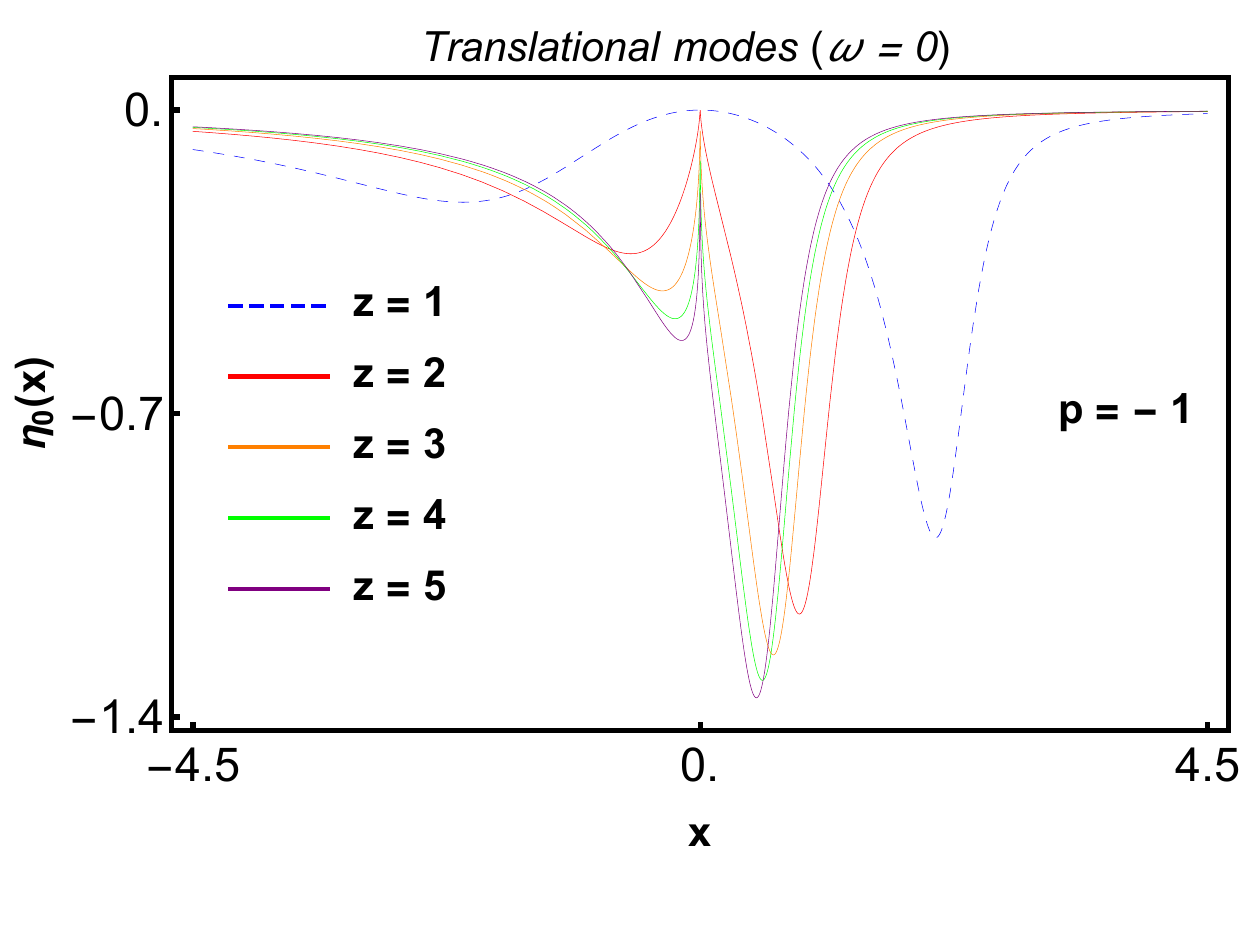}}\hfill
  \caption{Translational modes $\eta_0(x)$ vs. $x$ for right-handed asymmetry (i.e., $p=-1$) and varying $z$. In all cases, we adopt $\lambda=\nu=q=1$.}  \label{fig16}
\end{figure}

\begin{figure}[!ht]
  \centering
  \subfigure[$\eta_0(x)$ associated with asymmetric double-kink solutions.]{\includegraphics[height=5.5cm,width=5.4cm]{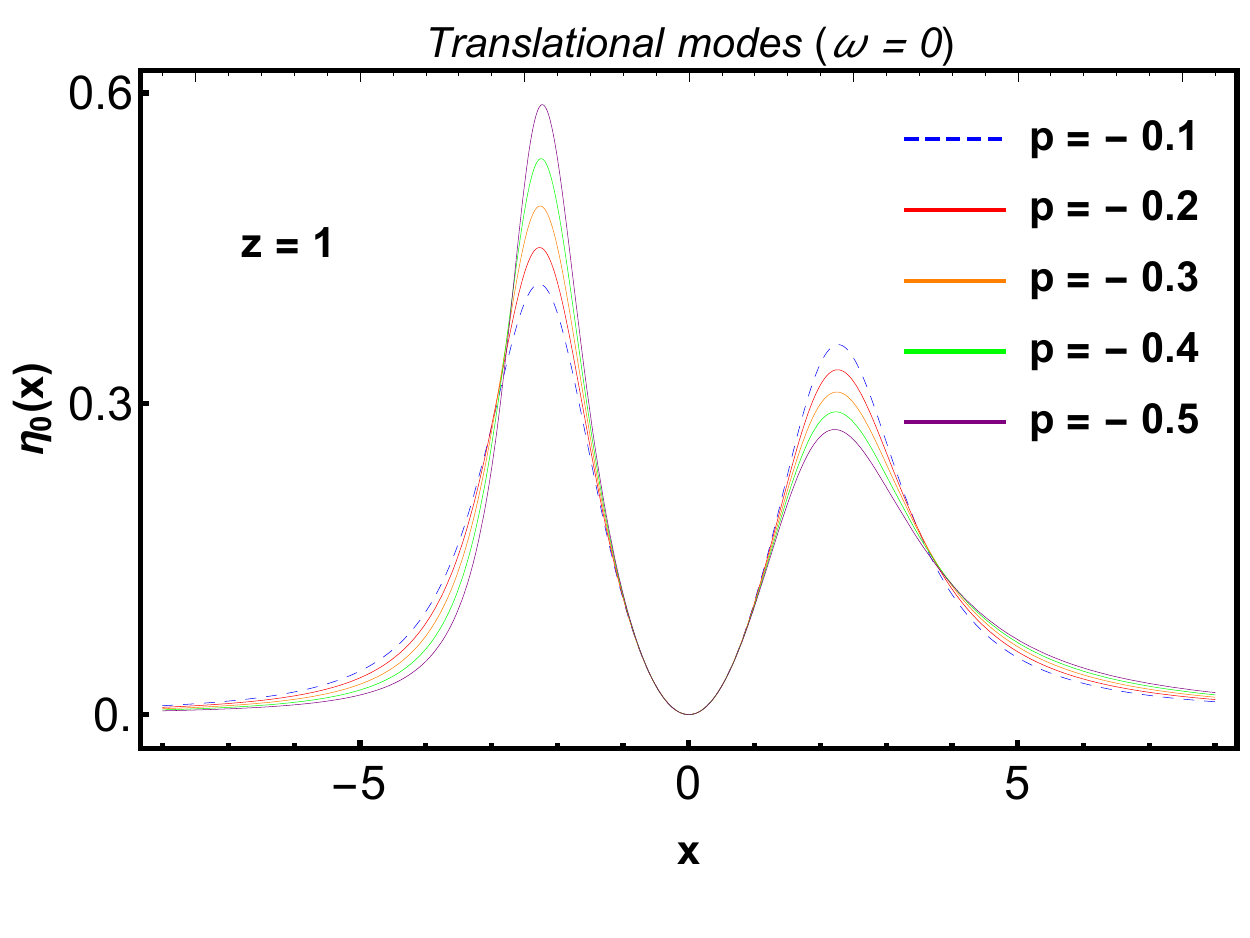}}\hspace{1cm}
   \subfigure[$\eta_0(x)$ associated with asymmetric double-antikink solutions.]{\includegraphics[height=5.5cm,width=5.4cm]{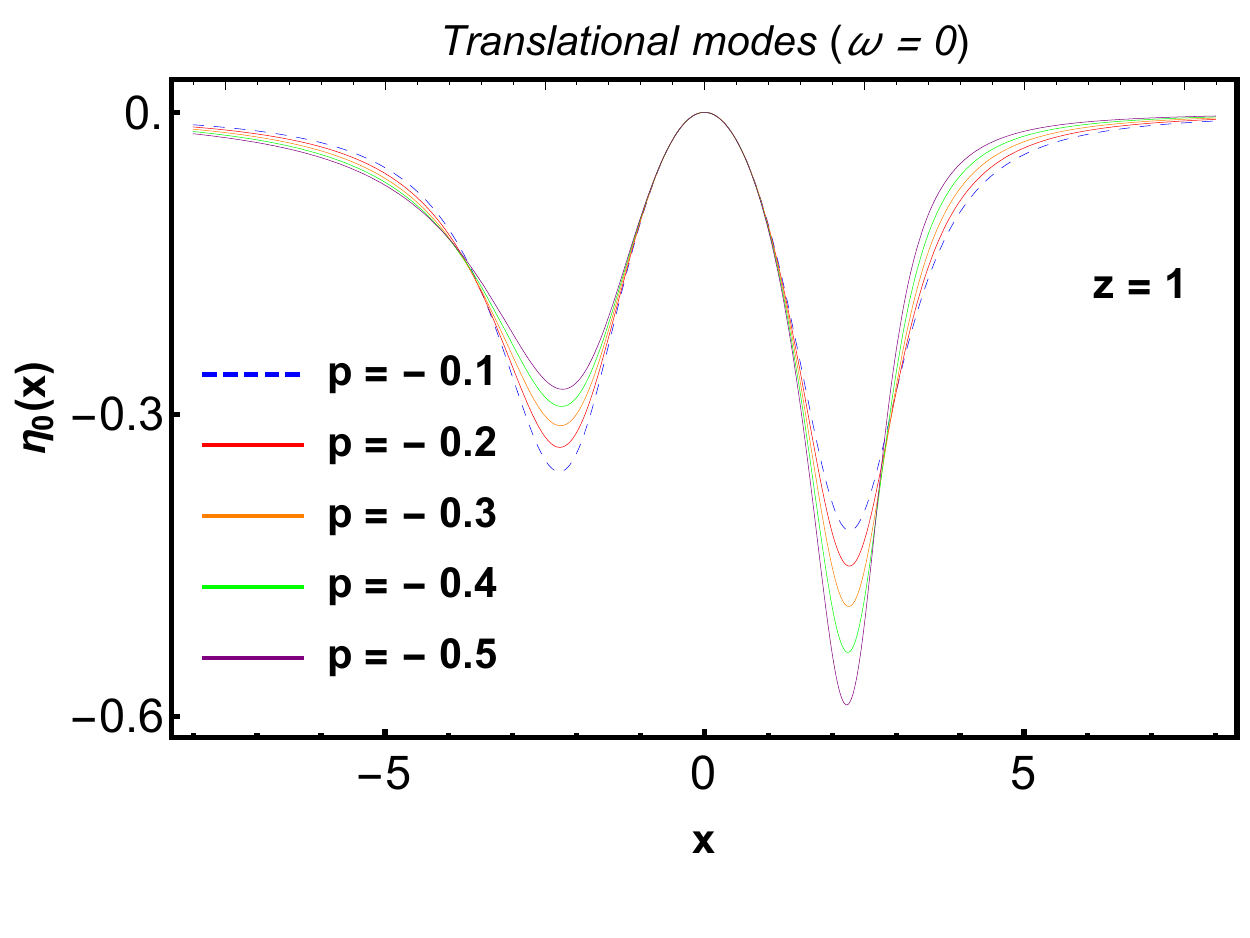}}\hfill
  \caption{Translational modes $\eta_0(x)$ vs. $x$ for right-handed asymmetry keeping $z=1$. In all cases, we consider $\lambda=\nu=q=1$.}  \label{fig17}
\end{figure}

\begin{figure}[!ht]
  \centering
  \subfigure[$\eta_0(x)$ associated with asymmetric double-kink solutions.]{\includegraphics[height=5.5cm,width=5.4cm]{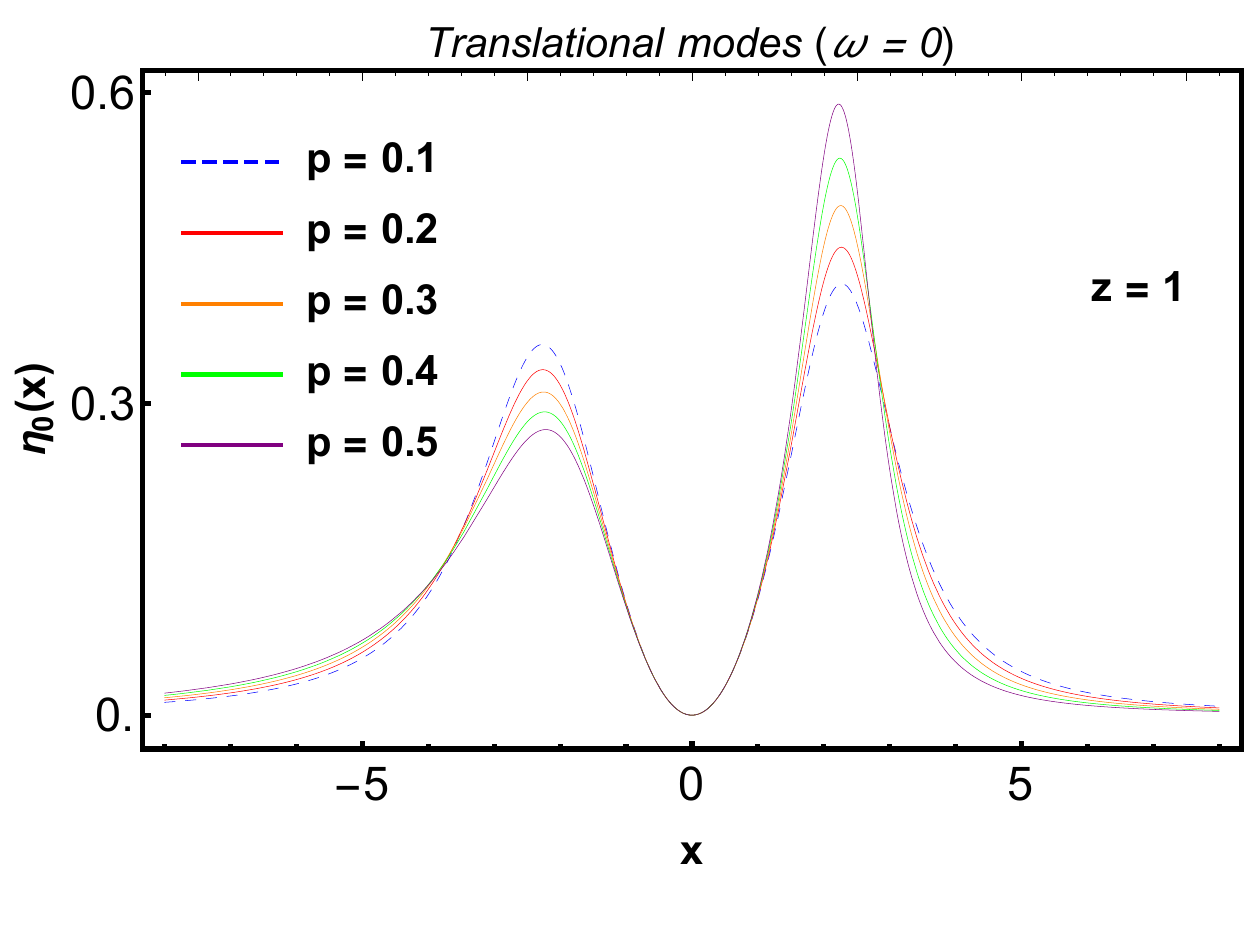}}\hspace{1cm}
   \subfigure[$\eta_0(x)$ associated with asymmetric double-antikink solutions.]{\includegraphics[height=5.5cm,width=5.4cm]{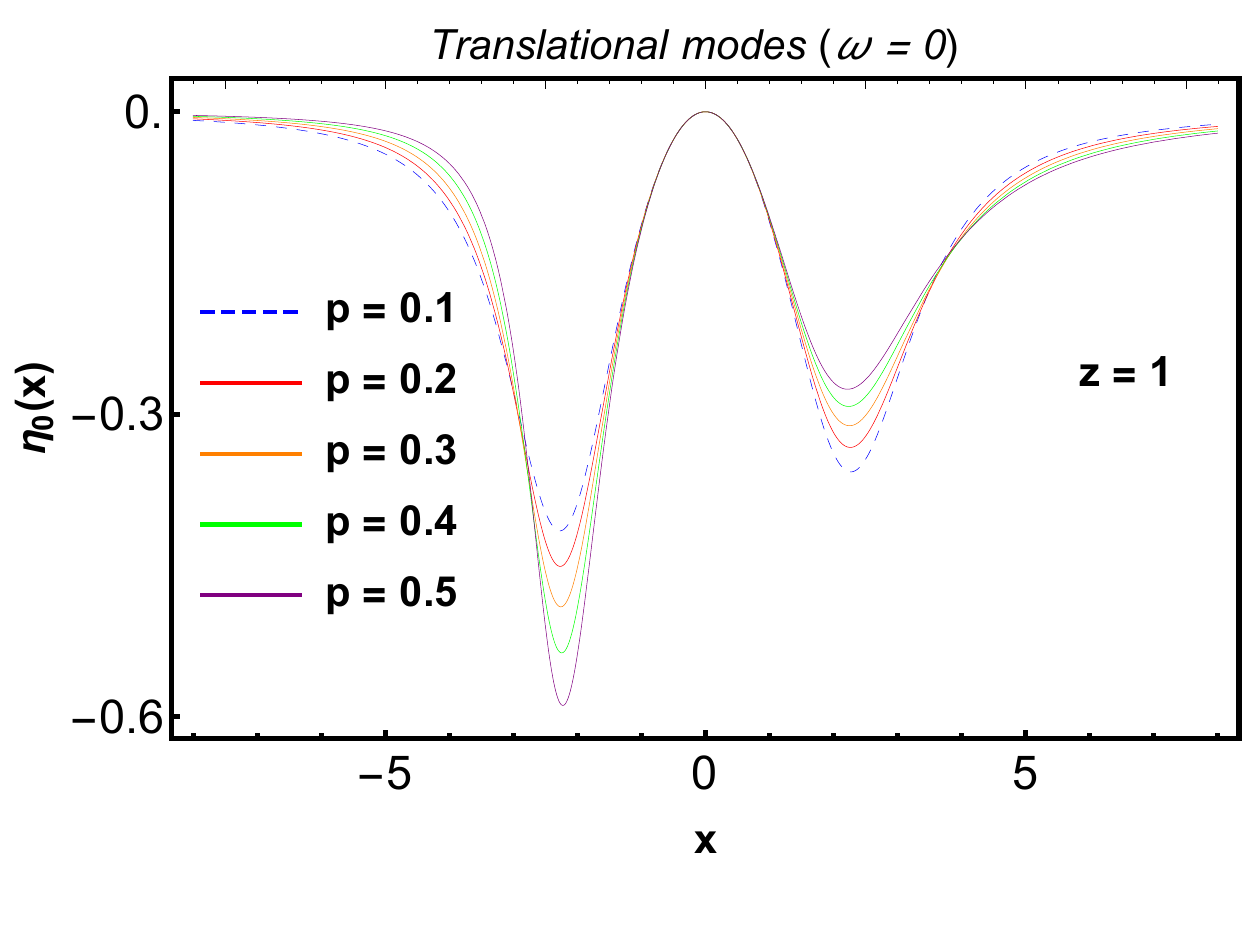}}\hfill
  \caption{Translational modes $\eta_0(x)$ vs. $x$ for left-handed asymmetry keeping $z=1$. In all cases, one assumes $\lambda=\nu=q=1$.}  \label{fig18}
\end{figure}

Assuming the Dirac delta-like behavior of the effective stability potential [Eq. \eqref{Eq29}], we conclude, by numerical inspection, the existence of at least a self-state, i.e., the translational mode. Thus, we now proceed to analyze the translational modes of our theory. To accomplish this purpose, let us adopt the numerical approach introduced in section \ref{sec2a}, in conjunction with Eq. \eqref{Eq25}. Through this approach, we expose the respective results in Figs. \ref{fig14}–\ref{fig18}(b).

Naturally, in the case without asymmetry ($p = 0$) and with deformation leading to the emergence of double-kink/antikink ($0< z < \infty$), the translational mode exhibits a profile with two symmetric local maxima around the center of the structure, i.e., at $x = 0$ (Fig. \ref{fig14}). This zero-mode profile confirms that the fields are double-kink and double-antikink configurations. Furthermore, the symmetric profile associated with the kink/antikink solutions indicates that, for genuine double-kink/antikink structures, there is no energy cost associated with their spatial translation \cite{Lima3}.

Meanwhile, asymmetry arises in the theory when $p \neq 0$ [see Figs. \ref{fig15}(a)–\ref{fig18}(b)]. In this framework, the asymmetry affects the translational modes, rendering them asymmetric in the double-kink/antikink. Resultantly, translating the structures to the right entails a different energy cost than translating them to the left due to the asymmetric profile of the potential $V(\phi)$.

Finally, it is worth noting that in models with deformation ($0 < z < \infty$) and asymmetry ($p \neq 0$), the translational mode satisfies the relation
\begin{align}
    \eta_{0}^{\text{(kink)}}(x)=-\eta_{0}^{\text{(antikink)}}(-x),
\end{align}
which confirms an asymmetry of the zero modes. Consequently, this result suggests that, in dynamical processes, such structures tend to be less stable and to scatter in opposite directions due to the dispersion of energy and momentum.

\section{Phenomenological consistency}\label{sec3}

The $\phi^4$ model describes a system in which spontaneous symmetry breaking is inherent. This symmetry breaking allows for the emergence of domain walls, i.e., field configurations that interpolate between the vacua (or domains with different field values). In this framework, such domains may represent, for instance, materials with distinct spin orientations, as in ferromagnet materials, or regions with different polarization states in ferroelectric materials.

Our model, deformed by a hyperbolic generalizing function, introduces a new domain with a zero field value, indicating the absence of a particle at $x = 0$ and, additionally, asymmetries in one of the preexisting domains of the $\phi^4$ model\footnote{\revision{Generally speaking, our generalized hyperbolic model introduces an anisotropy into the theory through the generalizing function $f(\phi)$, which modifies the critical phase transition and the collective excitations, reproducing a behavior similar to that of a polymeric SSH chain.}}. This theory is compatible with the physical description of a non-dimerized atomic chain\footnote{A non-dimerized atom does not participate in the formation of a dimer, i.e., it is not bonded to another identical atom to form a diatomic molecule. Structurally, it corresponds to an atom isolated concerning this specific type of pairing, which can have significant implications for the organization and physicochemical properties of the material.}, as such models preserve the occurrence of multi-kink-like domain walls.

\begin{figure}[!ht]
  \centering
  \includegraphics[height=3.5cm,width=8cm]{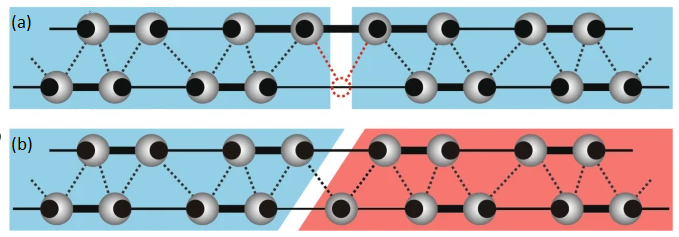}
  \caption{\revision{Phenomenological framework consistent with the theory \cite{Jeong}.}}  \label{fig19}
\end{figure}

However, through materials engineering, it is possible to eliminate the absence of the non-dimerized atom, resulting in the deformation of the domain wall at $x = 0$. By this means, atomic models arise in material engineering [Figs. \ref{fig19}(a) and \ref{fig19}(b)], without a single vacancy, become consistent with the domain walls generated by our generalizing function. Thus, one can interpret the domain with $\phi = 0$ as representing a vacancy that vanishes in the limit $z \to \infty$. Meanwhile, the asymmetry arises from the different domain configurations in distinct materials, as exemplified by the blue and red regions in Figs. \ref{fig9}(a) and \ref{fig19}(b), respectively.

In Figs. \ref{fig19}(a) and \ref{fig19}(b), we present a system that resembles our theoretical model. In this case, the red circles and dashed lines represent a missing atom and its interchain coupling \cite{Jeong,Huda}. Non-dimerized atoms in the chains are indicated by black (gray) circles pointing in opposite directions \cite{Jeong,Huda}. This domain wall structure contains a single non-dimerized vacancy located at the origin ($x = 0$).

The proposed theoretical model, together with the adopted phenomenological framework, describes double domain walls analogous to interacting Su–Schrieffer–Heeger (SSH) chains, connected through zigzag interchain couplings (represented by the dashed lines in Fig. \ref{fig19}). These couplings establish an energetic link between the asymmetric double-kink structures. Each SSH chain, in turn, undergoes a Charge Density Wave (CDW) transition, characterized by a doubling of the crystal structure’s period through Peierls dimerization\footnote{Peierls dimerization is a phenomenon in solid-state physics that occurs in one-dimensional electron systems along a periodic atomic chain. It demonstrates that a one-dimensional chain with an even number of atoms and one electron per atom is unstable in its uniformly spaced configuration and tends to undergo a spontaneous distortion.}. This transition is fundamental to the emergence of topological configurations and the nontrivial electronic properties associated with the domain walls.

However, the domain wall structure previously shown in Figs. \ref{fig19}(a) features a single vacancy when non-dimerized. Moreover, its characteristic zigzag interchain coupling also removes this absence from the crystal lattice of the polymer shown in Fig. \ref{fig19}(b). Additionally, the domain wall is not topological unless a constraint once the upper SSH chains in Fig. \ref{fig19}(a) and the lower ones in Fig. \ref{fig19}(b) are topologically trivial. To address these imperfections, we modify the domain wall by including the previously missing atom along a tilted cut in the polymer [Fig. \ref{fig19}(b)], which induces multiple asymmetric double-kink-like domain walls. In this way, our physical–mathematical results become consistent with the dynamics of a polyacetylene polymer chain
\footnote{Polyacetylene consists of a linear chain of carbon atoms with alternating single and double bonds, i.e., $-C=C-C=C-C=C-$. This structure admits two dimerization configurations (alternating bond patterns), which are energetically degenerate.}.

Thus, one highlights that the theoretical results presented may have relevant phenomenological implications, such as enhanced charge transport efficiency. Thereby, we have direct physical applications, e.g., consequences for electrical conduction in organic polymers, given that the solitonic profiles obtained are capable of carrying conserved charge, a nontrivial phenomenon from a physical standpoint. Therefore, these results provide a foundation for applications in low-dimensional systems and may impact emerging fields such as the development of new topological materials.

\section{Summary and conclusions}\label{sec4}

In this article, we performed a study on the mechanisms of deformation and asymmetry in domain walls within the framework of the modified $\phi^4$ theory. To accomplish our purpose, we adopted a non-canonical outline in which the kinetic term is $ \frac {f(\phi)}{2} \partial_\mu \phi\, \partial^\mu \phi$, leading to a generalized formulation\footnote{One can rescale the generalized theory into standard effective theory through the transformation $S \xleftrightarrow{\bar{\Phi}\leftrightarrow\int^{\phi}\,d\tilde{\phi} f^{1/2}(\tilde{\phi})} S_{\text{eff}}$.} governed by the action
\begin{align}
    S_{\text{eff}}=\int\,d^2x\,\left[\frac{1}{2}f(\phi)\,\partial_\mu\phi\,\partial^{\mu}\phi-\frac{\lambda}{2}(\nu^2-\phi^2)^2\right]
\end{align}
with
\begin{align}\label{fffc}
    f(\phi)=[\sinh(p\,\phi)+\cosh(p\,\phi)]^{-2}\sinh(q\,\phi^2)^{-\frac{1}{2z+1}}.
\end{align}
Here, the generalizing function $f(\phi)$, which distinguishes it from the usual $\phi^4$ theory, represents the hyperbolic contribution. That introduces deformation and asymmetry into our model. Physically, the generalizing function $f(\phi)$ allows for the emergence of new domains, i.e., a domain at $x = 0$, modeling new materials; for instance, the polymer chains discussed in section \ref{sec3}.

The choice of generalizing function adopted Eq. \eqref{fffc} leads to promising results regarding the behavior of the resulting topological structures, particularly by enabling the emergence of new classes of solutions, such as asymmetric and compact-like double-kink/antikink-like configurations. That occurs because, considering this generalizing function, new domains emerge violating the $\mathds{Z}_2$ symmetry\footnote{It is worth noting that in the limit $z \to \infty$ and $p \to 0$, the ordinary $\phi^4$ theory is recovered. Thus, we have the preservation of the $\mathds{Z}_2$ symmetry.}. %It is essential to highlight that this $\mathds{Z}_2$ symmetry breaking, in turn, implies the violation of translational invariance. 
Therefore, the asymmetric and deformed configurations become energetically less stable. Moreover, translating these structures in opposite directions results in different energetic costs. These results indicate that in a possible dynamical process involving the collision of asymmetric double-kinks (K) with asymmetric double-antikinks ($\tilde{K}$), i.e., a $K\tilde{K}$-like collision, the resulting behavior will differ from that observed in $\tilde{K}K$-like collisions. Such asymmetry in the collision process reinforces the nontrivial nature of the obtained solutions and their potential implications for the phenomenology of systems with broken symmetries.

Finally, we remarked that the results presented here are equivalent to the phenomenological outcomes observed in SSH polymer chains with zigzag interchain coupling in the absence of vacancies. Therefore, these findings are interesting due to their broad range of applications, such as theoretical proposals for new materials with enhanced charge transport efficiency. Thereby, the applications of this theory extend to electrical conduction and polarization in organic polymers, graphene, and other topological materials, where multi-soliton (i.e., multiply solitonic) structures are ubiquitous.

\section*{ACKNOWLEDGMENT}

The authors would like to express their sincere gratitude to the Conselho Nacional de Desenvolvimento Cient\'{i}fico e Tecnol\'{o}gico (CNPq) and Funda\c{c}\~{a}o de Amparo \'{a} Pesquisa do Estado de S\~{a}o Paulo (FAPESP) for their valuable support. F. C. E. Lima is supported, respectively, for grants No. 2025/05176-7 (FAPESP) and 171048/2023-7 (CNPq). C. A. S. Almeida is suported for grants No. 309553/2021-0 (CNPq)

\section*{CONFLICTS OF INTEREST/COMPETING INTEREST}

The authors declared that there is no conflict of interest in this manuscript. 

\section*{DATA AVAILABILITY}

No data was used for the research described in this article.

\end{document}